\documentclass[reprint,superscriptaddress,amsmath,amssymb,aps,prx]{revtex4-1}

\usepackage{graphicx}
\usepackage{dcolumn}
\usepackage{bm}
\usepackage{graphicx}
\usepackage{hyperref}
\usepackage{bbm}
\usepackage{xcolor}

\hypersetup{
  colorlinks,
  citecolor=magenta,
  linkcolor=magenta,
  urlcolor=magenta}

\def\be{\begin{equation}}
\def\ee{\end{equation}}

\def\nn{\nonumber}

\def\ra{\rangle}

\begin{document}
\title{Quantum resolution limit of long-baseline imaging using distributed entanglement}
\author{Isack Padilla}
\email{iacpad0795@arizona.edu}
\affiliation{College of Optical Sciences, University of Arizona, Tucson AZ 85721}

\author{Aqil Sajjad}
\email{asajjad@umd.edu}

\affiliation{College of Optical Sciences, University of Arizona, Tucson AZ 85721}
\affiliation{Department of Electrical and Computer Engineering, University of Maryland, College Park MD 20742}

\author{Babak N. Saif}
\email{babak.n.saif@nasa.gov}
\affiliation{NASA Goddard Space Flight Center, 8800 Greenbelt Rd, Greenbelt, MD 20771, USA}

\author{Saikat Guha}
\email{saikat@umd.edu}
\affiliation{College of Optical Sciences, University of Arizona, Tucson AZ 85721}
\affiliation{Department of Electrical and Computer Engineering, University of Maryland, College Park MD 20742}

\begin{abstract}
It has been shown that shared entanglement between two telescope sites can in principle be used to localize a point source by mimicking the standard phase-scanning interferometer, but without physically bringing the light from the distant telescopes together. In this paper, we show that a receiver that employs spatial-mode sorting at each telescope site, combined with pre-shared entanglement and local quantum operations can be used to mimic the most general multimode interferometer acting on light collected from the telescopes. As an example application to a quantitative passive-imaging problem, we show that the quantum-limited precision of estimating the angular separation between two stars can be attained by an instantiation of the aforesaid entanglement based receiver. We discuss how this entanglement assisted strategy can be used to achieve the quantum-limited precision of any complex quantitative imaging task involving any number of telescopes. We provide a blueprint of this general receiver that involves quantum transduction of starlight into quantum memory banks and spatial mode sorters deployed at each telescope site, and measurements that include optical detection as well as qubit gates and measurements on the quantum memories. We discuss the relative Fisher-information contributions of local mode sorting at telescope sites vis-a-vis distributed entanglement-assisted interferometry, to the overall quantum-limited information about the scene, based on the ratio of the baseline distance to the individual telescope diameter.
\end{abstract}

\maketitle

\section{Introduction}
In recent years, novel techniques inspired by quantum information theory have emerged in the field of super-resolution imaging, a field of study aimed at increasing the resolution of an imaging system by overcoming the Rayleigh criterion present in diffraction-limited systems, described in~\cite{LordRayleigh1879}.
One very promising technique in this regard is spatial-mode demultiplexing (SPADE), which involves spatial mode sorting of the light entering the imaging system prior to detection. The information-bearing light is thus measured in a nontrivial spatial mode basis instead of direct detection of its raw intensity profile on a standard imaging screen that amounts to measuring the incoming photons in the position (or a `pixel mode') basis. This approach has received considerable attention in recent years following Tsang {\em et al.}'s seminal paper on the problem of estimating the separation between two equally bright incoherent point sources~\cite{Tsang2016b}, at the ultimate precision allowed for by quantum estimation theory. Using the classical and quantum Cram\'er-Rao bounds, they show that while this problem suffers from Rayleigh's curse when direct detection is employed, a measurement in the image-plane Hermite-Gauss (HG) basis when we have a Gaussian point spread function (PSF), can bypass this limit and attain the optimal estimation precision allowed by the laws of physics, which is given by the so-called quantum Fisher Information (QFI).
This result was generalized to arbitrary aperture shapes by Kerviche \textit{et al.}~\cite{kerviche2017} and Rehacek \textit{et al.}~\cite{Rehacek2016}
who independently showed that an optimal SPADE basis for the two equally bright point separation problem can be generated by taking derivatives of increasing orders of the point spread function (PSF) and then performing a Gram-Schmidt orthogonalization process---henceforth termed the PSF-adapted (PAD) basis spatial mode basis. The PAD basis for a hard-aperture pupil in one dimension comprise of the Sinc-Bessel modes, for example.
More theory on this has been developed in~\cite{Tsang2016b, Ang2016, tsang2017,Tsang2019,Prasad2018}, such as extending this to two and three dimensions.
Other works have considered estimation of multiple parameters in a scene~\cite{Prasad2019,Rehacek2017b,rehacek2018,Prasad_2020,Bisketzi2019}, size estimation of extended objects~\cite{Zachary2019, Prasad2020B}, moment estimation of arbitrary scenes~\cite{Tsang2019,Tsang2019a,Zhou2019}, object discrimination~\cite{Lu2018,Huang2021,Grace2021c}, adaptively estimating the locations and brightnesses of several point sources of light~\cite{Bao2021,Matlin2022,Lee2022}, finding the precise location of an extended object or an array of point sources~\cite{Sajjad2021}, or estimating different parameters in two stages such as first finding the centroid (i.e., average position) and then the separation between two points~\cite{Tsang2016b, Grace2020c}.

These developments have also inspired some works on the application of quantum information theory techniques to the use of multiple optical receivers or apertures to perform long-baseline interferometry~\cite{Cosmo2020, Wang2021, Bojer2022, Sajjad2023}.
Such a collection of telescopes roughly emulates a single giant telescope with a size equal to the largest distance (baseline) between two nodes of the array, and its working principle consists of interferometrically exploiting the path-length difference of light arriving at two different observation points~\cite{Campbell1987,Lawson1999,Monnier2003}. 
In such a receiver, information can be extracted in two ways in principle. There is the information contained in the path difference between the light collected at different sites, and then there is also spatial resolution residing in the multi-spatial-mode light collected by each telescope. While most of the works in this field from a quantum information theory point of view have focused on the former alone and mostly taken individual receivers to be small, both effects have been considered in full detail in~\cite{Sajjad2023}, as the first such treatment. They discuss three broad sets of approaches for combining the light collected by different telescopes. First, when the telescopes are close to, e.g., within a few meters of, one another, we can use at each site a carved out piece of a giant parabola spanning the entire baseline, and focus the light on to a common imaging screen where it can be measured, e.g., by a single SPADE whose modal basis is matched to the problem at hand and the PSF of the effective compound aperture system. For larger distances where this is not feasible, we can use a spatial mode sorter at each site, couple the individual local first-few orders of information-rich orthogonal spatial modes into single mode fibers (SMFs), bring them to a central location, process them in a linear interferometer, and measure the outputs with shot-noise-limited photon detectors.
This is mathematically equivalent to carrying out a SPADE measurement on a (hypothetical) giant image plane spanning the entire baseline~\cite{Sajjad2023} discussed above. A related equivalent possibility is to employ multi-spatial-mode `light pipes' or vacuum beam guides~\cite{Huang2023} that carry the entire multi-mode light from each telescope to a central location, feed them into a linear interferometer made of bulk elements (e.g., beam splitters) that preserve the multi-spatial-mode information, and measure the outputs of the interferometer after spatial mode sorting at each output. One could use the light-pipe based interferometric receiver as above, but use bucket detectors at each output of the interferometer rather than spatial-mode-resolved detection. Such a strategy would reap the benefit of the baseline, but not extract scene information contained in local high-order spatial modes. 

Both approaches to baseline interferometry, i.e., using either single mode fibers or light pipes, become challenging to realize for baselines beyond a few hundred meters. Losses in single-mode fibers (for the fiber approach) and accurately transporting the multimode light without distortion and loss (for the light pipe approach) become prohibitive, not to mention the physical impracticality of laying light pipes over large geographical distances through oceans, mountains and other forms of complicated terrain.
For larger distances, to attain the fundamental resolution precision afforded by a long baseline system, the alternative is to rely on the (future) {\em quantum internet}---built using quantum repeaters and satellite-assisted links---that delivers error-corrected high-fidelity entanglement at high rates among quantum-memory banks deployed at the telescope sites, combined with local spatial-mode-resolved light collection and local interferometric measurements at each telescope site.

The simplest example that brings to light the aforesaid issues of quantum limit of a long baseline system, while quantifying the information contributions from mode sorting at each site vis-a-vis entanglement based interferometry, is to consider a two telescope set up for the problem of estimating the angular separation between two stars. At each telescope location, we collect $K$ orthogonal spatial modes with the modes indexed $q=0,\ldots, K-1$ and couple each into single mode fibers (SMFs), e.g., using a free-space-to-fiber coupled multi-plane light conversion (MPLC) mode sorter. Then, for each mode index $q$, we bring the respective SMFs from the two sites to a central location and interfere them in a 50-50 beam splitter, followed by detecting both outputs (of each of the $K$ beamsplitters), where the outputs of the $q$-th beam splitter are the sum and difference of the $q$-th spatial mode signals from the two telescopes. It has been shown in Ref.~\cite{Sajjad2023} that in the large $K$ limit, this pairwise interferometric scheme approaches the quantum Fisher information (QFI) limit of the precision for estimating the separation between two uniformly bright point sources in one dimension using two telescopes.

In this paper, we first describe how such a pairwise interferometric receiver can be implemented non-locally using pre-shared entanglement in place of single mode fibers and beam splitters. We do this by leveraging and building upon: (1) the quantum protocols for long baseline interferometry proposed by~\cite{Gottesman2012}, (2) its extension to employ solid-state spin-based quantum memories proposed by~\cite{Khabiboulline2018}, and (3) modal imaging techniques for super-resolution with a single telescope~\cite{Tsang2016a}. We then show how our approach can be generalized for carrying out parameter estimation in any complex quantitative imaging problem~\cite{Lee2022,Grace2022,Sajjad2023} at the quantum precision limit corresponding to a long-baseline system with any number of telescopes, using pre-shared multi-site entanglement and local operations at each site. We do this by mimicking a photon measurement in any arbitrary linear combination of optical modes collected at the telescope sites by a series of pairwise two-mode beamsplitter operations realized nonlocally via bi-partite pre-shared entangled Bell pairs. 

In the most general case, let us say we have $n$ telescopes indexed $\alpha=0,\dots,n-1$ where local spatial modes indexed $q=0,\ldots, K-1$ are collected and sorted at each one of them. We bring these $K$ modes, coupled into single mode optical fibers, to a central location where they are combined in a linear interferometer in a desired unitary transformation, i.e., one with $nK$ input modes and $nK$ output modes.
Such a linear optical interferometer realizing an arbitrary unitary mode transformation can be created using a total of $nK(nK-1)/2$ Mach-Zehnder interferometers (MZIs)~\cite{Clements2016}. Since a Mach-Zehnder interferometer comprises two 50-50 beam splitters and two single-mode phase shifters, a total of $nK(nK-1)$ 50-50 beam splitters and the same number of (tunable) single-mode phase shifters suffice to realize a general linear optical unitary. This means that if we can create the equivalence of the action of a 50-50 beam splitter (acting on a single-photon state spread across the two nonlocal modes on which the beamsplitter acts) by using pre-shared entanglement, in principle we can carry out $nK(nK-1)$ such nonlocal beamsplitters along with the same number of (local) single mode phases to reproduce the effect of an arbitrary beam splitter circuit consuming up to $nK(nK-1)$ pre-shared Bell pairs between quantum memories held at pairs of telescope sites, but without ever co-locating light from the $n$ telescopes. In this paper, we show how to realize the building block of the above argument---a two-input nonlocal beamsplitter acting on a photon spread across the two nonlocal modes---by first loading the photon mode at each site into a quantum memory~\cite{Khabiboulline2018} along with local quantum gates acting on those memories, and a preshared Bell pair among the two sites, via a sequence of local quantum gates.



\begin{figure*}
    \centering
    \includegraphics[scale=0.4]{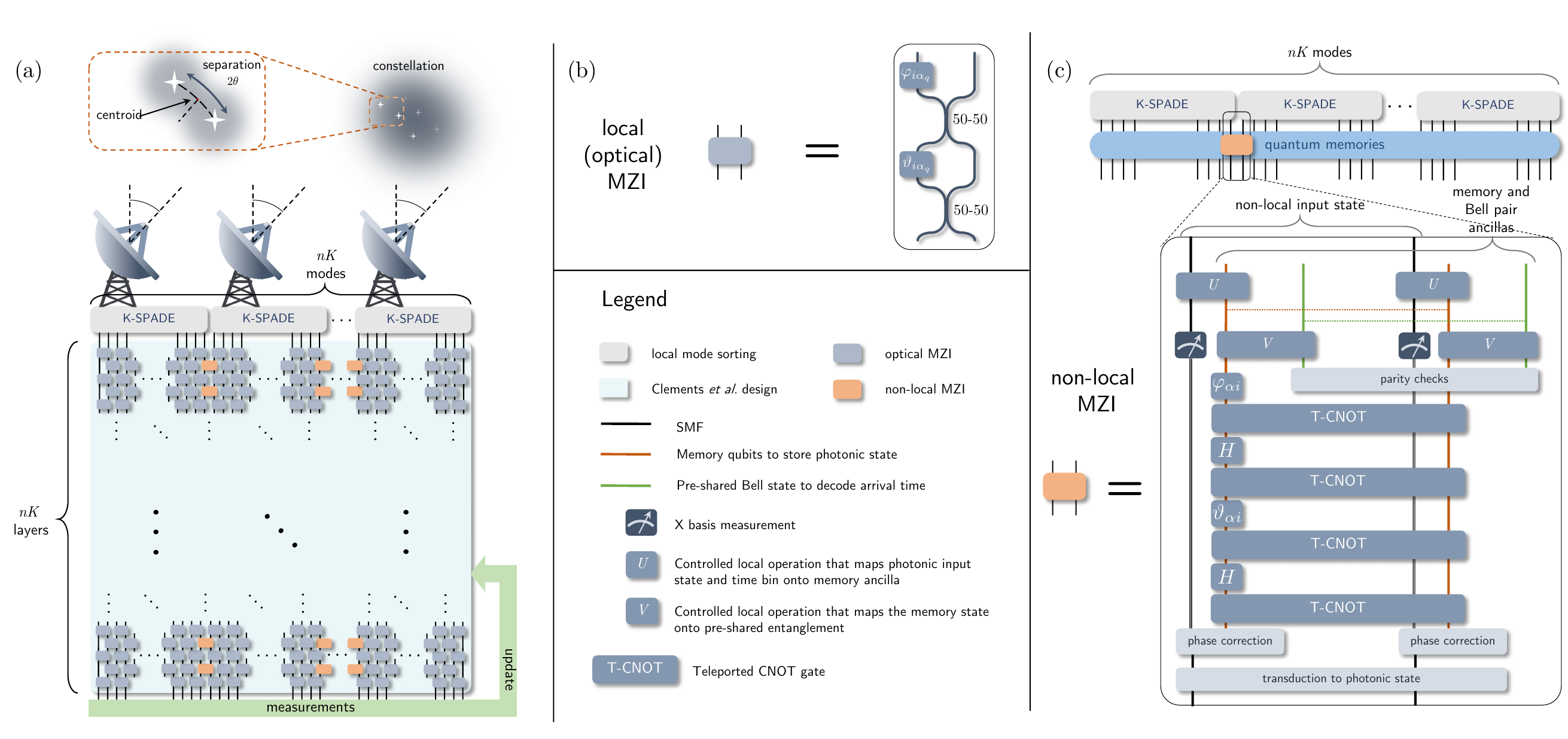}
    \caption{Overview of the system: (a) Light from a stellar scene is being collected at $n$ telescope sites. Each telescope employs a $K$-mode SPADE front-end that decomposes the incoming optical field into $K$ orthogonal spatial modes of a chosen basis. The sorted modes, across all the telescope sites, is fed into a general $nK$-mode linear-optical processor~\cite{Clements2016}, i.e., an array of Mach-Zehnder interferometers (MZIs) with two tunable phases $\vartheta_{\alpha i}$ and $\varphi_{\alpha i}$ in each, with $i \in \{1,…,nK\}$, $\alpha\in \{0,…,n-1\}$ and $q \in \{0,…,K-1\}$. In total, there are $nK (nK – 1)/2$ MZIs. (b) The gray boxes are an all optical MZI built with two 50-50 beamsplitters acting on two modes local at a telescope site. The orange boxes denote non-local MZIs, which act on a pair of modes located at two different telescope sites, which must be realized using pre-shared entanglement. (c) The non-local MZI (orange box) is realized by first transferring the joint two-mode optical state (black lines) onto a pair of atomic qubits (red lines) via the unitaries $U$ at each site followed by measuring off the photonic modes, then applying at each site two-qubit unitary gates $V$ between the aforesaid qubit and one half of a pre-shared ancilla Bell state (green line)---which maps the state of the memories to the ancilla Bell state imprinting on it the photon's arrival time. This arrival time is then retrieved by a parity-check measurement on the Bell states, followed by a sequence of local single-qubit gates and teleported CNOT gates~\cite{Chou2018}, and transducing the information-bearing memory qubits at each site back into the single-rail photonic qubit form (so that they can participate in subsequent telescope-site-local all-optical interferometry as prescribed by the task-specific global quantum-optimum measurement).}
    \label{fig1}
\end{figure*}

Therefore, local spatial-mode sorting at each telescope site, locally-applied single-mode phases and entanglement-assisted nonlocal realizations of two-mode 50-50 beamsplitters (see Fig. \ref{fig1}) can be used to compile the most general nonlocal multimode linear-optical pre-detection interferometry, thus helping attain quantum-limited resolution for quantitative imaging problems discussed above, limited only by the geometry of, and the light collected by the $n$-telescope baseline.

Before closing this section, we should mention how this paper fits into the overall picture in terms of existing proposals on the use of shared entanglement for baseline astronomy~\cite{Gottesman2012,Tang2016,Khabiboulline2018,Khabiboulline2019, Chen2023, Czupryniak2023}.
In the pioneering protocol for entanglement-based interferometry proposed by Gottesman \textit{et al.}, they consider the problem of estimating the phase difference between a pair of optical modes collected at the two telescopes sites carrying one incoming photon.
Say we have $n=2$ sites where we re-label telescope sites $A$ and $B$ for $\alpha=0$ and $\alpha=1$, respectively. A photon arriving from a star is in the state $\left(|1_A, 0_B\rangle +e^{i\theta} |0_A, 1_B\rangle\right)/\sqrt{2}$,
where $|1_A, 0_B\rangle$ is the state for the photon arriving at telescope $A$, and $|0_A, 1_B\rangle$ if it arrives at location $B$, with $\theta$ being a phase difference arising from the path difference for the photon between the two telescopes. This phase will naturally depend on the position of the star and the distance between the two telescopes, and therefore, in concrete terms, estimating $\theta$ amounts to determining the unknown location of a star emitting monochromatic light. In traditional baseline interferometry, we can estimate $\theta$ by bringing the light to a common location through `light pipes', put a tunable phase delay in one of these two signals, feed the two lightpipes into a 50-50 beam splitter, and measure the output intensities. As the tunable phase delay matches the equivalent optical phase shift of $\theta$, one of the two output intensities becomes zero due to fully destructive interference, and the other output contains all the light. Gottesman \textit{et al}.'s proposal mimics this technique by using pre-shared Bell states in the optical `single-rail' basis, i.e., $(|0,1\rangle + |1,0\rangle)/\sqrt{2}$, where $|1\rangle$ denotes a single-photon Fock state of a mode and $|0\rangle$ is vacuum, which can be generated by splitting a single photon in a 50-50 beamsplitter, distributed between the telescopes sites (reliably, say via quantum repeaters), local interferometric measurements at each telescope site, and a tunable phase at one telescope site. The main limitation of this scheme is that this entanglement must be provided at an unfeasible rate (at the modes-per-second corresponding to the optical bandwidth of the `science photons' being collected) with the technology available today, and that known repeater architectures---including all-photonic ones~\cite{Pant2017}---do not deliver photonic Bell states in the single-rail basis suitable for the aforesaid direct local interferometric measurements. They would deliver Bell states in the `dual-rail' photonic-qubit basis, converting which to single rail is not easy. Alternative approaches were proposed~\cite{Khabiboulline2018,Khabiboulline2019,Czu2022} for the aforesaid phase estimation utilizing atomic spin based quantum memories in cavities, assisted with a binary-code logarithm compression of the quantum state of one science photon spread across $M$ temporal modes into $\log_2(M)$ qubit memories, thereby resulting in an exponential reduction in the pre-shared Bell-state resource requirements compared to the previous proposal in~\cite{Gottesman2012}, as well as lending itself to a natural compatibility with atomic-qubit-based quantum repeater architectures~\cite{Guha2015,Dhara2021,Dhara2023,Lee2022a}.

While these contributions have nicely laid down the ground work for entanglement based interferometry, estimation of the phase difference between an incoming science photon between two telescope sites allows us only to estimate the position of a single point object. One could argue that with repeated use of this protocol, one could image any general scene. However, a suite of recent work on quantitative passive imaging problems shows that single-point-source localization is not a good proxy for imaging more complex scene features, and that the most photon efficient schemes---ones whose performance attains respective {\em quantum limits} of those quantitative imaging problems---must involve multimode, potentially adaptive, interferometry prior to detection~\cite{Lee2022,Grace2022}. In order to carry out more complex quantitative imaging and/or parameter estimation tasks, we need to be able to create an entanglement based equivalent of a multiple-telescope interferometric receiver where all the light received at different sites is brought through single mode fibers or `light pipes' to a central location, and fed into a linear interferometer whose outputs are then measured for the number of photon clicks~\cite{Sajjad2023}.

The paper is organized as follows. In Section~\ref{physical-set-up-section}, we explain our problem set up with two telescopes and two uniformly bright point sources, and summarize the main results of the paper. In Section~\ref{quantum-state-section}, we write down the joint quantum state of the incoming information-bearing light. We then dive into our nonlocal entanglement-assisted measurement protocol in Section~\ref{sec:encoding}, where the first step is to encode the state of the incoming photons at each telescope site on to memory qubits with the above-mentioned logarithmic compression idea. In Section~\ref{decoding-section}, we explain the decoding step where we carry out the measurements to determine the state of an incoming photon in terms of the combinations of the spatial modes collected at both sites. In Section~\ref{sec:quantumlimit_twosource}, we discuss how this approach attains the quantum Fisher-information (QFI) limit of estimation precision for the uniformly bright two-point separation problem. Section~\ref{sec:generalproblem} discusses how our protocol generalizes to any arbitrary measurement with any number of telescopes. In Section~\ref{sec:conclusions}, we conclude the paper with discussions pertaining to future work.
 
\section{Problem set up and main results}
\label{physical-set-up-section}

\begin{figure}
    \centering
    \includegraphics[scale=0.5]{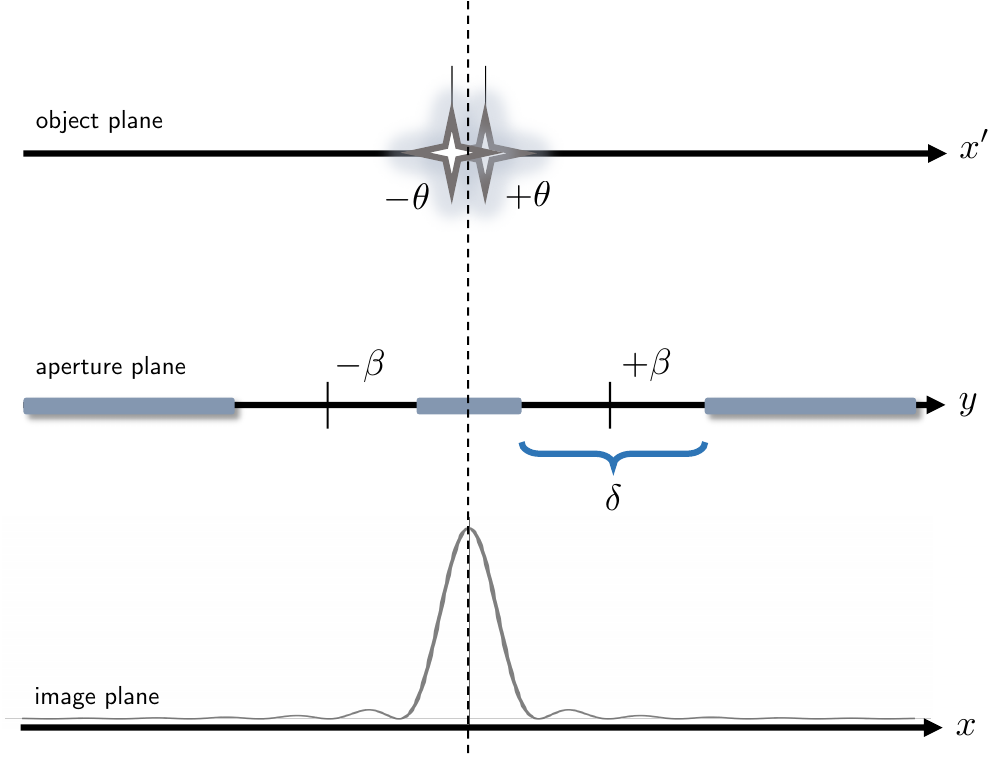}
    \caption{Schematic showing light originating at an object plane, propagating until reaching the aperture plane, with $n=2$ hard apertures each of size $\delta$, and finally propagation to an `image plane', where the light could be either detected with a pixelated photon detector array, or be fed into a SPADE followed by photon detection on the sorted modes~\footnote{For a long baseline system, where $\beta \gg \delta$, the image plane shown (and the `image-plane' fields and quantum states considered in the paper) is a hypothetical image plane that is a Fourier transform away from the (hypothetical) compound aperture plane comprising the two hard apertures separated by a distance, and hence `sees' the compound aperture's PSF, which is a centered sinc function times a cosine. In reality, each telescope employs a local (e.g., heterodyne) detection of the locally-collected optical field, or transmits the locally-collected field in a light pipe or optical fiber to a common site where light from different telescopes are brought into an interferometer.}.}
    \label{fig2}
\end{figure}

We consider the problem of estimating the half-separation $\theta$ between two equally-bright incoherently-radiating quasi-monochromatic point sources located at angular positions $\pm\theta$ (see Fig. \ref{fig2}). We will assume that a sufficiently-accurate prior knowledge of the centroid of the two sources has been obtained by  direct detection using a small fraction of the incoming photons during the integration time~\cite{Tsang2016b, Sajjad2021, Grace2020c}. Our imaging system consists of a two-telescope array labeled $\alpha\in\{A,B\}$, centered at $\pm \beta$, each of diameter $\delta$ (see Fig.~\ref{fig2}). The number of temporal modes $M$ of the incoming light contained within an integration time of $T$ (seconds) is: $M \approx WT$, where $W$ (Hz) is the optical bandwidth around the center frequency. We will consider imaging sources emitting light in optical frequencies (from UV to IR), in which regime the incoming light can be modeled as a weak thermal field. We will take the mean photon number per temporal mode $\epsilon$, in the incoming light collected by the two telescopes together, to be much less than one. We consider an integration time window $T$ commensurate with $M \approx 1/\epsilon$, such that there is roughly one photon but no more that arrives within that time window across both telescopes. We will adopt the compressive cavity-mediated light-matter interaction scheme from Ref.~\cite{Khabiboulline2018} to load the quantum state of $1$ photon spanning $M$ temporal modes, into $\log_2(M)$-qubit quantum memory registers located at each telescope site. The problem we consider is to estimate $\theta$ as accurately as possible---per the quantum-estimation-theory limit~\cite{Helstrom1976}---given a total of photons $N$ collected across both telescope sites.

Ref.~\cite{Sajjad2023} considered this problem, and showed that the Quantum Fisher Information (QFI) of $\theta$ (the inverse of which gives a tight bound on the minimum-permissible variance in estimation of $\theta$ when $N$ is large) is given by:
\begin{equation}
{\mathcal K} = \frac{4 \pi^2 N}{3 \sigma^2} \left(3r^2 + 1\right),
\end{equation}
where $r = 2\beta/\delta$ and $\sigma = 2\pi/\delta$ is the Rayleigh separation in wavelength-normalized unity-magnification-ratio units. Ref.~\cite{Sajjad2023} also showed that---provided the light collected by both apertures could be made available to be (processed and) detected in a common location with no loss---a receiver structure that achieves the QFI would employ spatial mode sorters at each aperture to sort sufficiently many (say $K \ge 1$) mutually-orthogonal spatial modes in the PAD bases (discussed in previous section) of each aperture, followed by projecting the two-mode state of the light collected---in each temporal mode, across both apertures in the $q$-th spatial mode, for each $q$, $0 \le q \le K-1$---into the symmetric or anti-symmetric mutually-orthogonal two-mode single-rail photonic-qubit Bell states:
\begin{align}
\left|\phi_{AB,q} ^\pm\right\rangle =\frac{1}{\sqrt{2}}\left(
|1_{A_q},0_{B}\rangle \pm |0_{A},1_{B_q}\rangle\right),
\quad q=0,\ldots,\infty
   \label{phipm_original}
\end{align}
where, $|1_{A_q},0_{B}\rangle$ (and $|0_{A},1_{B_q}\rangle$, respectively) is the state of a single photon in spatial mode $q$ at site $A$ (and $B$, respectively). In Ref.~\cite{Sajjad2023}, it was shown how to perform this projective measurement by mixing the aforesaid two modes, for each $q$, in a 50:50 beamsplitter, followed by photon detection on both output modes, and depending upon which detector clicks, the two-mode state would be projected on to one of two states $\left|\phi_{AB,q} ^+\right\rangle$ or $\left|\phi_{AB,q} ^-\right\rangle$. 

The next section of this paper will show in explicit terms: (1) how the above QFI-attaining projective measurement can be realized without bringing light contained in the two spatial modes $A_q$ and $B_q$ physically to mix on a beamsplitter, but by employing pre-shared entanglement stored in registers of atomic qubit memories located at each telescope site, (2) how to incorporate the compressive photon-to-memory state transfer from Ref.~\cite{Khabiboulline2018}---to load the quantum state of $1$ photon in $M$ temporal modes, into $\log_2(M)$ atomic qubits---while still being able to realize the underlying QFI-optimal projection~\eqref{phipm_original} from Ref.~\cite{Sajjad2023} on the quantum state of the incoming photon field, and (3) how to generalize this for a nonlocal entanglement-mediated realization of the QFI-optimal measurement for any quantitative imaging problem for a multi-telescope long-baseline system.

\section{Quantum state of the collected light, the `science photons'}
\label{quantum-state-section}

In this Section, we will express the joint quantum state of the `science photons', i.e., the light from a stellar scene comprising a constellation of equally-bright stars, collected at both telescope sites, across all the locally-sorted spatial modes at each telescope site and across $M$ temporal modes. We will assume the weak thermal state model described earlier, with $M \approx 1/\epsilon$ where $\epsilon \ll 1$ is the mean (thermal) photon number per mode. We will do this in three steps. In the first subsection, we write the quantum state for a single temporal mode at just one telescope site spanning all spatial modes. In the second subsection, we will generalize this to the joint quantum state of the light collected at $n=2$ telescopes over all spatial modes. Finally, in the third subsection, we write down the full quantum state across both telescopes, across all spatial modes, and $M$ temporal modes.

\subsection{Quantum state for a single-aperture telescope}

In the weak-thermal-source model~\cite{Tsang2016b}, the density operator---for a constellation of equally-bright point sources located at locations $x_s$ (where $s$ is a discrete index over the point sources making up the constellation)---of the state of a single temporal mode (but spanning all the spatial-modal content) at one telescope, is given by:
\begin{align}
    \rho=(1-\epsilon)\rho_0 + \epsilon\rho_1,
\end{align}
where we have made the approximation that $\epsilon$ is small enough to warrant considering terms only up to linear in $\epsilon$ and ignore $O(\epsilon^2)$ and higher-order terms. Here, $\rho_0=|0\rangle\langle0|$ denotes the vacuum state across the collected spatial-mode span, and the single-photon state $\rho_1$ is a mixture of pure states of the incoming light corresponding to the one-photon states of the image-plane shifted-PSF-modes corresponding to each star in the constellation; labeled $s\in\{1,2\}$ for our two-point-source problem:
\begin{align}
    \rho_1 = \frac{1}{2}\sum_{s\in\{1,2\}}\left|\psi^{(s)}\right\rangle\left\langle\psi^{(s)}\right|.
\end{align}
In the single aperture case, we can express:
\begin{align}
\left|\psi^{(s)}\right\rangle=\int \text{d}x \psi(x-x_s) |x\rangle,
\label{psis}
\end{align}
where $x$ is the image-plane coordinate, $|x\rangle=a^\dagger(x)|0\rangle$ is the state of a single photon at position $x$ (i.e., a single-photon state of a delta-function `pixel mode' in the image plane), and $a^\dagger(x)$ the bosonic creation operator in the continuum limit which satisfies the canonical commutation relation $[a(x),a^\dagger(x')]=\delta(x-x')$~\cite{yuen1978}. Lastly, $\psi (x-x_s)$ is the (laterally-shifted) point-spread function (PSF) corresponding to a single aperture, with $x_s$ being the position of the $s$-th star in the constellation.

Now, consider sorting the light collected by the aperture of the telescope in a spatial mode basis described by orthonormal mode functions $\phi_q(x)$ with $q=0, 1, 2,\ldots$. Expanding the state $\left|\psi^{(s)}\right\rangle$ defined in Eq. \eqref{psis} into this spatial-mode basis gives us:
\begin{align}
\left|\psi^{(s)}\right\rangle=\sum_{q=0}^{\infty} \Gamma_q(x_s) |1_q\rangle,
\label{psisq}
\end{align}
where 
\begin{align}
|1_q\rangle=\int \text{d}x \phi_q(x) |x\rangle
\end{align}
is the state of a single photon in the mode $\phi_q(x)$ (and vacuum in all the other modes in the orthogonal complement of $\phi_q(x)$), and the coefficients $\Gamma_q (x_s)$ are the correlation functions between the PSF and the mode functions:
\begin{align}
\Gamma_q(x_s) & =\left\langle 1_q\left|\psi^{(s)}\right.\right\rangle \\ 
& =  \int \text{d}x \phi_q^* (x) \psi(x-x_s).
\end{align}
If one detects photons on the sorted modes, if a photon arrival is detected, the probability of detecting that photon to be in the $q$th mode is then given by: $p_q(x_s) = |\Gamma_q(x_s)|^2$.

To obtain the PSF, we need to take the Fourier transform of the aperture function $\widetilde\psi(y)$, where $y$ is the aperture-plane coordinate. We will consider a one-dimensional hard aperture function:
\begin{equation}
\widetilde\psi (y) = \text{rect}\left(y/\delta\right)/\sqrt{\delta},
\label{rect-aperture-function}
\end{equation} 
where $\delta$ is the size (diameter) of the aperture. The Fourier transform of this yields the PSF
\begin{align}
\psi(x) = \sqrt{\sigma}\,
\frac{\sin(\pi x/\sigma)}{\pi x},
\label{PSF-sinc-def}
\end{align}
where $\sigma = 2\pi/\delta$ is the Rayleigh separation.

While we could detect the collected field in any mode basis, one natural basis set---which we will refer to as the {\em PSF-adapted} (PAD) basis---can be constructed by taking the PSF as the fundamental mode, taking derivatives of it of increasing order and normalizing the resulting functions to obtain a set of linearly-independent (but non-orthogonal) mode functions, and finally applying a Gram-Schmidt (GS) orthogonalization process to the above set to obtain the PAD-basis mode functions. For the above hard aperture PSF, the PAD basis is the Sinc-Bessel (SB) basis, for which the mode functions are given by:
$\phi_q(x)\equiv \sqrt{(1+2q)/\sigma}j_q(\pi x/\sigma)$,
 where $j_q(\pi x/\sigma)$ are the spherical Bessel functions of the first kind~\cite{kerviche2017, Rehacek2017a}. The correlation functions for this basis are
\begin{align}
    \Gamma_q (x_s) = \sqrt{\sigma}\phi_q(x_s).
\label{correlation-function-hard-aperture}
\end{align}
It was in fact shown in~\cite{Sajjad2023} that Eq.~\eqref{correlation-function-hard-aperture} exactly holds for any modal basis if we have the above said hard aperture PSF in 1 dimension. For a two-dimensional hard aperture, $\sqrt{\sigma}$ in Eq.~\eqref{correlation-function-hard-aperture} will be replaced by a slightly different constant that is proportional to the square root of the area of the hard apertures. We should however point out that the results and calculations given in this paper are mostly independent of the choice of the modal basis.

\subsection{Quantum state for the two telescope system}

We now consider two telescopes (apertures) at positions $y_A=-\beta$ and $y_B=\beta$ for which we will also use the notation $y_\alpha$ with $\alpha\in\{A,B\}$. We will mostly follow the framework described in~\cite{Sajjad2023}, which we now summarize, but with slightly modified notation and conventions.

When we have two apertures, the aperture function $\widetilde \psi(y)$ gets replaced by the normalized sum of two aperture functions centered at the two telescope locations {$(\widetilde \psi(y-y_A) + \widetilde \psi(y-y_\alpha))/\sqrt{2}$}.
Now, imagine a large (hypothetical) imaging screen behind the two apertures spanning the entire baseline so that the light entering the two hard apertures forms an image on it.
The PSF for this imaging system is the Fourier transform of this (compound) aperture function, just as the PSF for a single aperture is obtained by Fourier transforming the single-aperture function.
A lateral shift in a function translates into multiplication by a phase in the Fourier transform. Specifically, the Fourier transform of the shifted single-aperture function $\widetilde \psi(y-y_\alpha)$ is $e^{-iy_\alpha x}\psi(x)$, where $\psi(x)$ is the single-aperture PSF obtained by Fourier transforming the unshifted function $\widetilde \psi(y)$.
Applying this to the contributions from both apertures, the combined aperture function comes out to:
\begin{align}
\psi_{\text{2-ap}}(x) &= \sqrt{2} \cos(\beta x)\psi(x),
\label{psi-2ap-def}
\end{align}
that is, the PSF for a single unshifted aperture times a cosine term whose spatial frequency contains information about the baseline (separation) of the two-aperture system, with a $\sqrt{2}$ arising from the normalization.
For the special case of two hard, rect shaped apertures with the aperture function \eqref{rect-aperture-function} centered at positions $\pm\beta$ in the aperture plane, we obtain the compound PSF:
\begin{align}
\psi_{\text{2-ap, h}}(x) = \sqrt{2\sigma}
\cos(\beta x) \, \frac{\sin(\pi x/\sigma)}{\pi x},
\label{compound-psf-2-apertures-hard}
\end{align}
where the additional subscript (h) denotes `hard' apertures. We will refer back to this in the last section of the paper when we evaluate the quantum-limited performance of the measurement carried out by our protocol, limited only by this compound aperture.

The quantum state of a single photon originating from a point source at object-plane position $x_s$ and entering our two-aperture system is an equal linear superposition of the states corresponding to the starlight arriving at site $A$ and at site $B$:
\begin{align}
\left|\psi^{(s)}_{AB}\right\rangle 
&= \int \text{d}x \psi_{\text{2-ap}}(x) |x\rangle \nonumber \\
&= \frac{1}{\sqrt{2}} \left(\left|\psi_{A}^{(s)}\right\rangle + \left|\psi_{B}^{(s)}\right\rangle\right),
    \label{psiab}
\end{align}
where $|x\rangle$ is the position ket for a photon hitting an imaginary imaging screen spanning the entire baseline,
and $|\psi_A^{(s)}\rangle$ and $|\psi_B^{(s)}\rangle$
are the kets associated with the photon arriving at the two locations
\begin{align}
\left|\psi_\alpha^{(s)}\right\rangle
= \int \text{d}x e^{-i y_\alpha (x - x_s)} \psi (x - x_s) |x\rangle.
\quad \alpha\in\{A,B\}
    \label{psi1}
\end{align}

In the same spirit, the mode functions $\phi_q(x)$ also acquire the same phase factor as $\psi(x)$ when the aperture is shifted to a position $y_\alpha$.
This can be seen by thinking in terms of the aperture plane where $\phi_q(x)$ translates into its Fourier transform $\widetilde \phi_q(y)$.
If the aperture is shifted to a location $y_\alpha$ in the aperture plane, then the mode function becomes $\widetilde \phi_q(y-y_\alpha)$, which in the image plane  gives $\exp(-i y_\alpha x) \phi_q(x)$.
We then have the state
\begin{align}
|1_{A_q},0_{B}\rangle\equiv a_{A_q}^{\dagger}|0_A,0_B\rangle
= \int \text{d}x e^{-i y_A x}\phi_{q}(x) |x\rangle
\label{local-modes-site-A}
\end{align}
for a single photon in the (local) spatial-mode $q$ of the telescope at site $A$,
and
\begin{align}
|0_{A},1_{B_q}\rangle\equiv a_{B_q}^{\dagger}|0_A,0_B\rangle
= \int \text{d}x e^{-i y_B x}\phi_{q}(x) |x\rangle
\label{local-modes-site-B}
\end{align}
for a single photon in the (local) spatial-mode $q$ at site $B$.
Here {\small $|0_\alpha\rangle \equiv \bigotimes_{i=0}^{\infty}  |0_{\alpha_{i }}\rangle$}
is the vacuum in all the spatial modes at site $\alpha$,
in terms of which our position ket on an imaginary imaging screen spanning the entire baseline is
$|x\rangle = a^{\dagger}(x) |0_{A},0_{B}\rangle$, and {\small$a_{\alpha_q}^{\dagger}=\int \text{d}x e^{-i y_\alpha x}\phi_{q}(x)a^{\dagger}(x)$} is the creation operator for a photon in the $q$-th mode at site $\alpha$.

We will refer to $|1_{A_q},0_{B}\rangle$ and $|0_{A},1_{B_q}\rangle$ as the single-photon states of the respective $q$-th local (shifted PAD-basis) spatial mode of the telescopes at sites $A$ and $B$, respectively. The local mode function set of one aperture will be orthogonal to those of another one, which is easy to see in the $y$-plane (see Fig.~\ref{fig2}) considering the two local PAD-basis mode sets are mutually spatially non-overlapping (given the hard apertures). Overall, the local modes of all our apertures together provides a complete orthonormal modal basis for our multiple aperture system.

Having specified how the local mode functions and the PSF acquire phase factors when we shift an aperture from the origin to a position $y_\alpha$ in the aperture plane, we can expand the ket $|\psi^{(s)}_\alpha\rangle$ associated with a photon from a star at position $x_s$ arriving in aperture $\alpha$ in terms of its local modes. From \eqref{psi1}, \eqref{local-modes-site-A} and \eqref{local-modes-site-B}, we obtain
\begin{align}
\left\langle 1_{A_q},0_{B} \left| \psi^{(s)}_A\right.\right\rangle
&=e^{i y_A x_s}\Gamma_{q}(x_s) \nonumber, \,{\text{and}} \\
\left\langle 0_{A},1_{B_q}\left|\psi^{(s)}_B\right.\right\rangle
&=e^{i y_B x_s}\Gamma_{q}(x_s).
\label{coeff}
\end{align}

Using this result and expanding \eqref{psiab} in terms of the local modes, we obtain
\begin{align}
\left|\psi^{(s)}_{AB}\right\rangle =& \frac{1}{\sqrt{2}}\sum_{q=0}^{\infty} \Gamma_{q}(x_s) \left( e^{i y_A x_s} |1_{A_q},0_{B}\rangle + e^{i y_B x_s} |0_{A},1_{B_q}\rangle\right) \nonumber \\
=& \frac{1}{\sqrt{2}}\sum_{q=0}^{\infty} \Gamma_{q}(x_s)
\left( e^{-i \beta x_s} |1_{A_q},0_{B}\rangle + e^{i \beta x_s} |0_{A},1_{B_q}\rangle\right),
\label{psiab2}
\end{align}
where in the second line, we have plugged in $y_A=-\beta$ and $y_B=\beta$ for the respective telescope positions.

In a realistic physical set up, we will only be able to collect a finite number of spatial modes at each site. Assuming that we are collecting only the first $K$ local PAD-basis modes, this results in the above state being projected on to the subspace limited to $0\leq q\leq K-1$,
\begin{align}
\left|\psi^{(s)}_{AB}\right\rangle 
\to \left|\psi^{(s)}_{AB, K}\right\rangle \equiv \frac{1}{\sqrt{2}}\sum_{q=0}^{K-1}\eta_q(x_s)&\left( e^{-i \beta x_s} |1_{A_q},0_{B}\rangle\right. \nonumber \\ & \left.+ e^{i \beta x_s} |0_{A},1_{B_q}\rangle\right),
\label{psiab3}
\end{align}
where 
\begin{align}
\eta_q(x_s) \equiv \frac{\Gamma_q(x_s)}{\sqrt{\sum_{l=0}^{K-1} \Gamma_l ^2(x_s)}}
\label{eta-definition}
\end{align}
are the new coefficients obtained from dividing $\Gamma^2_q(x_s)$ by the normalization factor {\small$\sqrt{\sum_{l=0}^{K-1} \Gamma_l ^2(x_s)}$}.
However, for a sufficiently large $K$, this factor will approach 1, and we recover $\eta_q(x_s) \to \Gamma_q(x_s)$. In the same spirit, the photon flux will also change if we only collect the first $K$ modes, being redefined as
\begin{align}
N_K \equiv N\sum_{l=0}^{K-1} \Gamma_l ^2(x_s),
\label{N_k-def}
\end{align}
which will approach $N$ for a sufficiently large $K$ for which {\small$\sum_{l=0}^{K-1} \Gamma_l ^2(x_s)$} will approach unity.

Our goal is to use shared entanglement to measure each temporal mode of the incoming photon field in the pairwise bases~\cite{Sajjad2023}:
\begin{align}
\left|\phi_{AB,q} ^\pm\right\rangle =\frac{1}{\sqrt{2}}\left(
|1_{A_q},0_{B}\rangle \pm |0_{A},1_{B_q}\rangle\right),
    \label{phipm}
\end{align}
while ensuring that the measurement strategy heralds the temporal modes that contain a photon (like the original physical strategy from~\cite{Sajjad2023} of pairwise beamsplitters followed by photon detection would). The projection of \eqref{psiab3} onto this measurement basis yields probabilities:
\begin{align}
    p_q^+ = \eta^2 _q (x_s) \cos^2(\beta x_s)
    \label{pplus1}
\end{align}
and
\begin{align}
    p_q^- = \eta^2 _q (x_s) \sin^2(\beta x_s)
    \label{pminus1}
\end{align}
for the two outcomes, which are the probabilities we want to replicate with the entanglement-based protocol to be presented in this paper. In the limit where we collect a large number of spatial modes i.e. $K\to\infty$, these approach $\Gamma^2 _q (x_s) \cos^2(\beta x_s)$ and $\Gamma^2 _q (x_s) \sin^2(\beta x_s)$, respectively,
which are the probabilities obtained in~\cite{Sajjad2023}.

Having spelled out the ket for a photon originating from a given point source, we can write down the joint density operator describing the weak incoming signal, arriving at sites $A$ and $B$,
\begin{align}
    \rho_{AB}=(1-\epsilon)\rho_{0,AB} + \epsilon\rho_{1,AB},
\end{align}
where $\rho_{0,AB}=|0_A,0_B\rangle\langle 0_A, 0_B|$ is the vacuum state across both locations across their entire local spatial-mode spans, which we defined above, and the one-photon state $\rho_{1,AB}$ for the two aperture system is a mixture of the one-photon states associated with the two point sources:
\begin{align}
    \rho_{1,AB} = \frac{1}{2}\sum_{s\in\{1,2\}}\left|\psi^{(s)}_{AB, K}\right\rangle\left\langle\psi^{(s)}_{AB, K}\right|,
\end{align}
where $|\psi^{(s)}_{AB, K}\rangle$ was given in \eqref{psiab3}, and the factor of $\frac{1}{2}$ arises from equal brightness of the two point sources.

\subsection{Quantum state of the collected light for two telescopes over $M$ temporal modes}

As described in Section~\ref{physical-set-up-section}, we will be interested not only in a nonlocal realization of the  projection described in Eq.\eqref{phipm} on each temporal mode (while heralding the presence of a photon in each temporal mode), but we also wish to do that in a way so as to allow for the compressive loading of $1$ photon in $M$ temporal modes into $\log_2(M)$ qubit memories per spatial mode at each site
(to minimize the shared-entanglement resource consumption). We consider a measurement time interval $T$ (seconds) such that the number of temporal modes $M = WT \approx 1/\epsilon$, where $\epsilon$ is the mean photon number per temporal mode (spanning both telescopes) and $W$ (Hz) is the optical bandwidth of the source (after any spectral filtering performed by the telescope). The state of the incoming light is $\rho^{\otimes M}_{AB}$, with $m\in\{1,\ldots,M\}$ indexing the temporal modes in the time interval. By computing this tensor product (see Appendix \ref{A1} for a detailed derivation), and keeping only terms linear in $\epsilon$ (because the choice of $T$, hence $M$, ensures at most one photon during time $T$), we obtain:
\begin{align}
    \rho^{\otimes M}_{AB} \approx (1-M\epsilon)\rho_{0,AB}^{\otimes M} + \epsilon\sum_{m=1}^M \rho_{1,AB,m},
\label{rhoabM}
\end{align}  
where $\rho_{1,AB,m}$ is a tensor product of $M$ states of which $M-1$ are vacuum and the remaining one is a one-photon state $\rho_{1,AB}$ in the $m$-th position (of the tensor product):
\begin{subequations}
 \begin{eqnarray}
\rho_{1,AB,m}&=& \rho_{0,AB}^{\otimes (m-1)}\otimes\rho_{1,AB}\otimes\rho_{0,AB}^{\otimes (M-m)} \\
&=& \rho_{0,AB}^{\otimes (m-1)}\otimes\frac{1}{2}\sum_{s\in\{1,2\}}\left|\psi^{(s)}_{AB, K}\right\rangle\left\langle\psi^{(s)}_{AB, K}\right| \nonumber \\
&&\otimes\rho_{0,AB}^{\otimes (M-m)} 
\\ &=&\frac{1}{2}\sum_{s\in\{1,2\}}\left|\psi^{(s)}_{AB,K, m}\right\rangle\left\langle \psi^{(s)}_{AB,K,m} \right|. 
\label{rho1abmc}
\end{eqnarray}   
\end{subequations}
The ket in Eq.~\eqref{rho1abmc} is that of a single photon from star $s$ arriving in temporal mode indexed $m$ (in the time interval $T$), and is defined as:
\begin{widetext}
\begin{subequations}
\begin{align}
\left|\psi^{(s)}_{AB,K, m} \right\rangle &\equiv |0_A,0_B\rangle^{\otimes(m-1)}\otimes \left|\psi^{(s)}_{AB, K}\right\rangle \otimes|0_A,0_B\rangle^{\otimes(M-m)}\\
&=\dfrac{1}{\sqrt{2}}\sum_{q=0}^{K-1} \eta_q (x_s) \left(e^{i\beta x_s}|0_A,0_B\rangle^{\otimes(m-1)}\otimes|0_A,1_{B_q}\rangle\otimes|0_A,0_B\rangle^{\otimes(M-m)}\right. \nonumber\\
    & \left. \;\;\;\;+ e^{-i\beta x_s}|0_A,0_B\rangle^{\otimes(m-1)}\otimes|1_{A_q},0_{B}\rangle\otimes|0_A,0_B\rangle^{\otimes(M-m)} \right) \\
&=\dfrac{1}{\sqrt{2}}\sum_{q=0}^{K-1} \eta_q (x_s) \left(e^{i\beta x_s}\left|\widetilde0_{A},\widetilde1_{B_{mq}}\right\rangle+ e^{-i\beta x_s}\left|\widetilde1_{A_{mq}},\widetilde0_{B}\right\rangle  \right),
\end{align}  
\end{subequations}
\end{widetext}
where $|\psi^{(s)}_{AB, K}\rangle$ was given in \eqref{psiab3}, and
\begin{align}
 \left|\widetilde 0_\alpha\right\rangle \equiv \bigotimes_{i=0}^{K-1} \bigotimes_{j=1}^M |0_{\alpha_{ji}}\rangle,
\quad \alpha\in\{A,B\},
 \label{zeroket}
\end{align}
and
\begin{align}
\left|\widetilde 1_{\alpha_{mq}}\right\rangle = a_{\alpha_{mq}} ^\dagger \left|\widetilde 0_\alpha\right\rangle,
\quad \alpha\in\{A,B\},
\end{align}
are the states corresponding to vacuum across the $KM$ modes and one photon in the spatio-temporal mode $(m,q)$, respectively.
The creation operator $a_{\alpha_{mq}}^\dagger$ produces a photon in the $(m,q)$-th spatio-temporal mode at telescope site $\alpha$.

Since we are only considering zero and single photon states across the $M$ temporal modes, for the subsequent sections when we discuss photon-memory interactions for loading the science photons into quantum memories, we will often find it useful to think of the state associated with each spatio-temporal mode of the collected optical field at a given telescope site as a single rail photonic qubit, i.e., a qubit encoded in the span of $|0\rangle$ (vacuum state of that spatio-temporal mode) and $|1\rangle$ (the state of $1$ photon in that spatio-temporal mode).

In the subsequent Sections, we address the different stages towards entanglement-assisted (nonlocal) interferometry of the incoming photon field collected at the two telescopes across the $KM$ spatio-temporal modes described above. First we go through the encoding process in which the information in the science photons is mapped onto qubits of a quantum memory register in the fashion of~\cite{Khabiboulline2018,Khabiboulline2019}. In the Section of decoding, we will show how the state of the memories are mapped onto ancilla Bell pairs for (teleported) transfer of the information from one of the aperture sites to the other, followed by a final measurement sequence that effectively realizes the desired projection~\eqref{phipm} on the original (optical) state of all the $KM$ mode pairs at the two telescope sites.

\section{Encoding photonic quantum state into quantum memories}\label{sec:encoding}

The encoding step involves transferring the state of the photons to a quantum memory state 
described by sets of systems $\overline{A}$ and $\overline{B}$, labeled $\overline{\alpha}\in\{\overline{A},\overline{B}\}$, which start as a register, for a given spatial mode, of $\overline{M}=\log_2(M+1)$ qubits initialized in the state $|0\rangle$.
We introduce indices $k\in\{1,\dots,\overline{M}\}$, $i\in\{0,\dots,K-1\}$ and $j\in\{1,\dots,M\}$ that account for the resources used throughout the protocol (memory qubits, CNOT gates and pre-shared Bell pairs, respectively), with the special case $i=q$ and $j=m$ corresponding to the spatio-temporal mode $(m, q)$ where the photon is present.
For both sites we write the state of the memory qubits as
\begin{align}
{\sigma}_{0}=\left|0_{\overline A},0_{\overline B}\right\rangle\left\langle 0_{\overline A},0_{\overline B}\right|,
\label{initial-memory-qubits}
\end{align}
where we have defined
\begin{align}
\left|0_{\overline\alpha}\right\rangle\equiv\bigotimes_{i=0}^{K-1}\bigotimes_{k=1}^{\overline{M}}|0_{\overline{\alpha}_{ki}}\rangle,
\quad \alpha\in\{A,B\},
\end{align}
so a total of $2K\overline{M}$ memory qubits are employed.

We then wish to load the photonic state on to the memory qubits. From hereon, when referring to one of the $KM$ optical modes, we will synonymously use {\em photonic qubit} to refer to the single-rail photonic qubit (of the respective mode). We first employ a collection of CNOT gates from the $KM$ photonic qubits to the $K{\bar M}$ memory qubits, such that a photonic excitation in the $i$th spatial mode of the $j$th temporal mode at a given telescope site ($\alpha$ or $\beta$) translates into excitation(s) in the binary representation of the integer $j$ into the memory qubits of the memory-register corresponding to spatial mode $i$ at that telescope location. These CNOT operations for site $\alpha$ can be denoted by $U_{\alpha_{ji},\overline{\alpha}_{ki}}$, where we identify the control and target qubits as the photonic qubit in the $i$-th spatial mode and $j$-th temporal mode (photonic qubit $\alpha_{ji}$) and the $k$-th qubit in the memory (memory qubit $\overline{\alpha}_{ki}$), respectively.
The full unitary operator acting on the state for all spatial and temporal modes at site $\alpha$ is:
\begin{align}
{U}_{\alpha\overline\alpha}=\bigotimes_{i=0}^{K-1}\bigotimes_{j=1}^{M}\bigotimes_{k=1}^{\overline{M}}{({U_{\alpha_{ji}\overline{\alpha}_{ki}}})^{w_{kj}}},
\quad \alpha\in\{A,B\}
\label{CNOT-at-each-site}
\end{align}
Here the symbol $w_{kj}$ represents the $k$th digit in the binary form of $j$; e.g., if we have a photon in $j=5$, its binary representation is the bit-string $1\,0\,1$, then $w_{15}=1$, $w_{25}=0$, $w_{35}=1$ and $w_{k'5}=0$ for $k'\in\{4,\ldots,\overline{M}\}$.

The action of the CNOTs at each site \eqref{CNOT-at-each-site} thus gives 
\begin{align}
{U}_{\alpha\overline\alpha}
|\widetilde1_{\alpha_{mq}}\rangle|0_{\overline \alpha}\rangle
= |\widetilde1_{\alpha_{mq}}\rangle|1_{\overline \alpha_{mq}}\rangle,
\quad \alpha\in\{A,B\}
\end{align}
where  
\begin{align}
|1_{\overline\alpha_{ m q}}\rangle
\equiv \bigotimes_{i=0}^{K-1}\bigotimes_{k=1}^{\overline{M}} X_{\overline\alpha_{i k}} ^{\delta_{i q} w_{km}} |0\rangle_{\overline{\alpha}_{ki}}
\end{align}
is the logical representation of the mapped excitation in the memory. Here $X_{\overline\alpha_{ik}}$ denotes the Pauli $X$ operator acting on the memory qubit of system $\overline\alpha$ corresponding to the $k$-th memory qubit of spatial mode $i$, flipping it from $|0\rangle$ into $|1\rangle$.

Now, the combined state of the photon and the blank memory is $\rho_{AB}^{\otimes M}\otimes\sigma_{0}$,
and the encoding is performed by the operator
$U\equiv {U}_{A \overline A}\otimes{U}_{B\overline B}$.
Then, the photon-memory state after applying the CNOTs is 
\begin{align}
    {\rho}_{AB\overline{AB}} &= 
U\rho_{AB}^{\otimes M}\otimes\sigma_{0}
    U^{\dagger} \\
    & = (1-M\epsilon)\rho_{0,AB}^{\otimes M}\otimes\sigma_{0}
    \nonumber \\ &\;\;\;\;+ \frac{\epsilon}{2}\sum_{\substack{m=1 \\ s\in\{1,2\}}}^M\left|\psi^{(s)}_{AB\overline{AB},m}\right\rangle\left\langle \psi^{(s)}_{AB\overline{AB},m} \right|,
\end{align}
with the following ket obtained after applying the compound CNOT unitary $U$ on the photon and memory states:
{\small
\begin{align}
    \left| \psi_{AB\overline{AB},m}^{(s)} \right\rangle \equiv &\dfrac{1}{\sqrt{2}}\sum_{q=0}^{K-1} \eta_q (x_s) \left(e^{i\beta x_s}\left|\widetilde0_{A},\widetilde1_{B_{mq}}\right\rangle\left|0_{\overline A},1_{\overline B_{mq}}\right\rangle \right. \nonumber\\
    & \left. + e^{-i\beta x_s}\left|\widetilde1_{A_{mq}},\widetilde0_{B}\right\rangle \left|1_{\overline A_{mq}},0_{\overline B}\right\rangle \right).
    \label{psiabab}
\end{align}}
The result from this mapping is thus an entangled state between the single-photon quantum state of the $2MK$ spatio-temporal modes of the collected field and the $2{\bar M}$ qubits of the two ${\bar M}$-qubit memory registers at the telescope sites. The next step is to disentangle the two and separate the photon from the memory. This is done by measuring the photonic qubits (systems $A$ and $B$) in the $X$ basis $|\pm\rangle=(|0\rangle\pm|1\rangle)/\sqrt 2$.
Such measurement of the photonic modes will introduce a conditional phase in the state of the memory qubits. One way we can deal with this is by expanding the photonic portions of~\eqref{psiabab} into a sum of four $X$ basis tensor product states. First note that going from the $Z$ basis to the $X$ basis for two modes yields the following:
 \begin{align}
    |0_{A_{ji}},0_{B_{ji}}\rangle &= \frac{1}{2}\left(|+_{A_{ji}} +_{B_{ji}}\rangle +|-_{A_{ji}}  -_{B_{ji}}\rangle \right.\nonumber \\ & \quad\left.  |+_{A_{ji}} -_{B_{ji}}\rangle+|-_{A_{ji}} +_{B_{ji}}\rangle\right),
\end{align}
\begin{align}
    |1_{A_{ji}},0_{B_{ji}}\rangle &= \frac{1}{2}\left(|+_{A_{ji}} +_{B_{ji}}\rangle-|-_{A_{ji}} -_{B_{ji}}\rangle \right.\nonumber \\ & \quad\left.  +|+_{A_{ji}} -_{B_{ji}}\rangle-|-_{A_{ji}} +_{B_{ji}}\rangle\right),
\end{align}
 \begin{align}
    |0_{A_{ji}},1_{B_{ji}}\rangle &= \frac{1}{2}\left(|+_{A_{ji}} +_{B_{ji}}\rangle-|-_{A_{ji}} -_{B_{ji}}\rangle \right.\nonumber \\ & \quad\left.  -|+_{A_{ji}} -_{B_{ji}}\rangle+|-_{A_{ji}} +_{B_{ji}}\rangle\right).
\end{align}
This allows us to re-write the photon-memory state \eqref{psiabab} in the form
\begin{widetext}
\begin{align}
    \left| \psi_{AB\overline{AB},m}^{(s)} \right\rangle = &   \frac{1}{\sqrt{2}}\sum_{q=0}^{K-1}\eta_q(x_s)\left|e_{AB,mq}\right\rangle\left(e^{i\beta x_s}\left|0_{\overline A},1_{\overline B_{mq}}\right\rangle
+e^{-i\beta x_s}\left|1_{\overline A_{mq}},0_{\overline B}\right\rangle\right)  \nonumber  
\\ &  \quad + \frac{1}{\sqrt{2}}\sum_{q=0}^{K-1}\eta_q(x_s)\left|o_{AB,mq}\right\rangle\left(e^{i\beta x_s}\left|0_{\overline A},1_{\overline B_{mq}}\right\rangle-e^{-i\beta x_s}\left|1_{\overline A_{mq}},0_{\overline B}\right\rangle\right)  ,
\label{photon-memory-state-even-odd-expansion}
\end{align}
\end{widetext}
where
\begin{align}
\left|e_{AB,mq}\right\rangle = \frac{1}{\sqrt 2}\left(\left|+_{A_{mq}},+_{B_{mq}}\right\rangle-\left|-_{A_{mq}},-_{B_{mq}}\right\rangle\right)
    \label{e-ab}
\end{align}
and
\begin{align}
    \left|o_{AB,mq}\right\rangle = \frac{1}{\sqrt 2}\left(\left|-_{A_{mq}},+_{B_{mq}}\right\rangle-\left|+_{A_{mq}},-_{B_{mq}}\right\rangle\right)
    \label{o-ab}
\end{align}
are the ``even" and ``odd" photonic states, respectively.
The kets $\left|\pm_{\alpha_{mq}}\right\rangle$ contained in \eqref{e-ab} and \eqref{o-ab} are the states in which all the photonic qubits are in $|0\rangle$ except the one corresponding to the spatio-temporal mode $(m, q)$, which is in $|\pm\rangle$, at site $\alpha$. More concisely, we can write such states as $|+_{\alpha_{mq}}\rangle
=  H_{\alpha_{mq}}|\widetilde 0_{\alpha}\rangle$ and $|-_{\alpha_{mq}}\rangle
=  Z_{\alpha_{mq}}H_{\alpha_{mq}}|\widetilde 0_{\alpha}\rangle$, where $H_{\alpha_{mq}}$ and $Z_{\alpha_{mq}}$ are the Hadamard and Pauli Z operators acting solely on spatio-temporal mode $(m, q)$ of \eqref{zeroket}, respectively.

Thus from \eqref{photon-memory-state-even-odd-expansion}, we see that measuring the photonic qubits in the $X$ basis dis-entangles the photon-memory system, but gives a conditional sign flip in the memory state depending on whether we obtain odd $(\pm,\pm)$ or even $(\pm,\mp)$ parity results for the qubits associated with the temporal and spatial modes $(m, q)$ at the two sites.
and the resulting memory state is
\begin{align}
\left|\chi_{\overline{AB},m}^{(s)}(\textbf{h}) \right\rangle = \dfrac{1}{\sqrt{2}}\sum_{q=0}^{K-1}\eta_q(x_s) \left(e^{i\beta x_s } \left|0_{\overline A},1_{\overline B_{mq}}\right\rangle  
    \right. \nonumber \\ \left.
    +f(\textbf{h}_{mq})e^{-i\beta x_s }\left|1_{\overline A_{mq}},0_{\overline B}\right\rangle\right).
    \label{memory-ket}
\end{align}
Here the conditional sign is given by
\begin{align}
f(\textbf{h}_{mq})=
\begin{cases}
+1 & \text{for } h_{A_{mq}}=\pm \text{ and } h_{B_{mq}}=\pm \\
-1 & \text{for } h_{A_{mq}}=\pm \text{ and } h_{B_{mq}}=\mp \\
\end{cases},
\end{align}
where $h_{A_{mq}}$ and $h_{B_{mq}}$
are the results of the $X$ basis measurements of the photonic qubits corresponding to the time bin $m$ and spatial mode $q$ at the two sites.

We should mention that at this point, we do not yet know the temporal and spatial mode $(m, q)$ of the incoming photon, and therefore cannot tell if $f(\textbf{h}_{mq})$ is $1$ or $-1$. Thus at this stage, we have to simply store the results of all the $X$ basis measurement results in a classical memory register for now. Once we go through the decoding stage and determine the temporal and spatial mode, then we will be able to look at these stored results and see if $f(\textbf{h}_{mq})$ is $1$ or $-1$, depending on whether the corresponding photonic qubit results have even or odd parity, respectively. Therefore, $h_{A_{mq}}$ and $h_{B_{mq}}$ are entries of a $2KM$-component array $h$ containing the results of all the $X$ basis measurement outcomes. 


The density matrix for the memory system is then
\begin{align}
\sigma_{\overline{AB}}(\textbf{h}) &= (1-M\epsilon)\sigma_{\overline{AB},0} \nonumber \\
& \quad + \frac{\epsilon}{2}\sum_{\substack{m=1 \\ s\in\{1,2\}}}^M \left| \chi_{\overline{AB},m}^{(s)}(\textbf{h})\right\rangle\left\langle \chi_{\overline{AB},m}^{(s)}(\textbf{h})\right|.
\label{sigmah}
\end{align}
The encoding of the arrival time is complete, nonetheless, we require the collapse of the superposition over index $q$ via a projective measurement, which is accomplished in the decoding step which we describe in the next section. The implementation of the encoding would require the interaction of the incoming photon with an atomic cavity, similar to the one described in Ref.~\cite{DuanKimble2004}. As for the measurement of a single-rail photonic qubit in the $X$ basis, it has been suggested in Ref.~\cite{Khabiboulline2019} that it could be approximated by mixing the photon with ancillary states in an interferometer, and in principle, the larger the interferometer, the closest we could be to a deterministic $X$-basis measurement. A more careful analysis of this proposal and an alternative method for an in-principle deterministic $X$ basis measurement on single-rail photonic qubits will be presented in future publications~\cite{Sajjad2025b, Sajjad2025c}.

\section{Decoding}
\label{decoding-section}

We now proceed with the decoding step where we reveal the time bins in which the photons arrive. The employed tool is an ancilla consisting of $K\overline{M}$ entangled pairs described by the sets of systems $ C$ and $ D$ .
This ancilla state has the form
\begin{align}
    |\Phi_{CD}\rangle=\bigotimes_{i=0}^{K-1}\bigotimes_{k=1}^{\overline{M}} \left| \phi^+_{C_{ki}D_{ki}} \right\rangle,
\end{align}
where $| \phi^+_{C_{ki}D_{ki}} \rangle=(|0_{C_{ki}},0_{D_{ki}}\rangle+|1_{C_{ki}},1_{D_{ki}}\rangle)/\sqrt{2}$.
We will then employ a circuit that flips some or all of these into
$| \phi^-_{C_{ki}D_{ki}} \rangle=(|0_{C_{ki}},0_{D_{ki}}\rangle-|1_{C_{ki}},1_{D_{ki}}\rangle)/\sqrt{2}$,
if some of the memory qubits are in $|1\rangle$ due to an incoming photon in the original photonic system.
The Bell states upon measurement in the $X$ basis, will yield `even' or `odd' parity results and reveal which of the Bell states underwent a flip and allow us to determine the spatio-temporal mode of the incoming photon, as we will shortly describe. This step is also known as `checking' the parity.

Systems $C_{ki}$ and $D_{ki}$ are sent to sites $A$ and $B$, respectively. The decoding step involves the application of Controlled $Z$ (CZ) gates where the state of the memory acts as the control qubit whereas the target qubit comes from the pre-shared entangled ancilla (note that all the $| \phi^+_{C_{ki}D_{ki}} \rangle$ pairs are identical and the index is used to keep track of which Bell pair is acted on by the CZ gates). The mapping is carried out with the operator defined as
\begin{align}
    V\equiv\bigotimes_{i=0}^{K-1}\bigotimes_{k=1}^{\overline{M}}\left(V_{\overline A_{ki}C_{ki}}\otimes V_{\overline B_{ki}D_{ki}}\right),
\end{align}
where the indices ($\overline A_{ki}$, $\overline B_{ki}$) and ($C_{ki}$, $D_{ki}$) represent the control and target qubits, respectively.
The effects of these gates are
{\small 
\begin{subequations}
    \begin{align}
    V_{\overline{A}_{ki}, C_{ki}} V_{\overline{B}_{ki}, D_{ki}} & \left|0_{\overline{A}_{ki}}, 1_{\overline{B}_{ki}}\right\rangle \left|\phi^+ _{C_{ki} D_{ki}}\right\rangle \nonumber\\ & = \left|0_{\overline{A}_{ki}}, 1_{\overline{B}_{ki}}\right\rangle \left|\phi^- _{C_{ki} D_{ki}}\right\rangle,
    \label{vv1}
\end{align}
\begin{align}
    V_{\overline{A}_{ki}, C_{ki}} V_{\overline{B}_{ki}, D_{ki}} & \left|1_{\overline{A}_{ki}}, 0_{\overline{B}_{ki}}\right\rangle \left|\phi^+ _{C_{ki} D_{ki}}\right\rangle \nonumber\\ & = \left|1_{\overline{A}_{ki}}, 0_{\overline{B}_{ki}}\right\rangle \left|\phi^- _{C_{ki} D_{ki}}\right\rangle,
    \label{vv2}
\end{align}
\begin{align}
    V_{\overline{A}_{ki}, C_{ki}} V_{\overline{B}_{ki}, D_{ki}} & \left|0_{\overline{A}_{ki}}, 0_{\overline{B}_{ki}}\right\rangle \left|\phi^+ _{C_{ki} D_{ki}}\right\rangle \nonumber\\ &= \left|0_{\overline{A}_{ki}}, 0_{\overline{B}_{ki}}\right\rangle \left|\phi^+ _{C_{ki} D_{ki}}\right\rangle,
    \label{vv3}
\end{align}
\begin{align}
    V_{\overline{A}_{ki}, C_{ki}} V_{\overline{B}_{ki}, D_{ki}} & \left|1_{\overline{A}_{ki}}, 1_{\overline{B}_{ki}}\right\rangle \left|\phi^+ _{C_{ki} D_{ki}}\right\rangle \nonumber\\ & = \left|1_{\overline{A}_{ki}}, 1_{\overline{B}_{ki}}\right\rangle \left|\phi^+ _{C_{ki} D_{ki}}\right\rangle,
    \label{vv4}
\end{align}
\end{subequations}}
i.e., for odd memory pairs, $ |\phi^+_{C_{ki}D_{ki}}\rangle$ transforms into $|\phi^-_{C_{ki}D_{ki}}\rangle=(|0_{C_{ki}},0_{D_{ki}}\rangle-|1_{C_{ki}},1_{D_{ki}}\rangle)/\sqrt{2}$, and even pairs leave the state $ |\phi^+_{C_{ki}D_{ki}} \rangle$ unchanged.
Thus with the application of the CZ operations from the memory state \eqref{memory-ket} to systems $C$ and $D$, we obtain 
\begin{widetext}
\begin{subequations}
\begin{align}
\left| \chi_{\overline A \overline B C D,m}^{(s)}(\textbf{h})\right\rangle
&= V\left| \chi_{\overline{AB},m}^{(s)}(\textbf{h}) \right\rangle |\Phi_{CD}\rangle \\
&= V \dfrac{1}{\sqrt{2}}\sum_{q=0}^{K-1}\eta_q(x_s) \left(e^{i\beta x_s } \left|0_{\overline A},1_{\overline B_{mq}}\right\rangle  
     +f(\textbf{h}_{mq})e^{-i\beta x_s }\left|1_{\overline A_{mq}},0_{\overline B}\right\rangle\right)\bigotimes_{i=0}^{K-1}\bigotimes_{k=1}^{\overline{M}} \left| \phi^+_{C_{ki}D_{ki}} \right\rangle \\
    &= \sum_{q=0}^{K-1}\eta_q(x_s)  \left|\chi_{\overline{AB},mq}^{(s)}(\textbf{h})\right\rangle |\Phi_{CD,mq}\rangle,
     \label{abcd}
\end{align}
\end{subequations}
\end{widetext}
where 
\begin{align}
\left|\chi_{\overline{AB},mq}^{(s)}(\textbf{h})\right\rangle=\dfrac{1}{\sqrt{2}} \left(e^{i\beta x_s } \left|0_{\overline A},1_{\overline B_{mq}}\right\rangle 
\right. \nonumber \\ \left.
     +f(\textbf{h}_{mq})e^{-i\beta x_s }\left|1_{\overline A_{mq}},0_{\overline B}\right\rangle\right)
\end{align}
is the superposition of memories corresponding to an excitation in modes $q$ and $m$, and
\begin{align} 
    \left|\Phi_{CD,mq}\right\rangle
&=\bigotimes_{i=0}^{K-1}\bigotimes_{k=1}^{\overline{M}} \left(Z_{C_{ki}} \right)^{\delta_{iq}w_{km}}\left|\phi^+_{C_{ki}D_{ki}}\right\rangle \nonumber \\ 
&=\bigotimes_{i=0}^{K-1}\bigotimes_{k=1}^{\overline{M}} \left(Z_{D_{ki}} \right)^{\delta_{iq}w_{km}}\left|\phi^+_{C_{ki}D_{ki}}\right\rangle 
\label{phi_Mq-def}
\end{align}
is the Bell state ancilla after the unitary $V$ has been applied. The Pauli operators $Z_{C_{ki}}$ and $Z_{D_{ki}}$
have the effect of flipping the sign of the relevant Bell pairs according to the rules \eqref{vv1}-\eqref{vv4} (by acting on qubit $(k, i)$ of system $C$ and $D$), respectively.
The power $\delta_{iq}$ means that this operator is acting only when $i=q$, and $w_{km}$ means it is happening only for the Bell pairs corresponding to the binary representation of the temporal mode number $m$.
In short, if we have an incoming photon in temporal mode $m$, then in each term of our state, the Bell pairs corresponding 
to the spatio-temporal mode $(m, q)$ undergo a flip, while the remaining Bell states are unflipped.
 On the other hand, the action of $V$ on the first term of \eqref{sigmah} associated with the photonic vacuum of systems $A$ and $B$, will leave all the Bell pairs unaffected: $V|0_{\overline A},0_{\overline B}\rangle|\Phi_{CD}\rangle=|0_{\overline A},0_{\overline B}\rangle|\Phi_{CD}\rangle$.

We now need to determine which Bell pairs flipped. For this, note that expressing the flipped and unflipped Bell states in the $X$ basis, we obtain
\begin{equation}
    \left|\phi^+_{C_{ki}D_{ki}}\right\rangle=\dfrac{1}{\sqrt{2}}\left(|+_{C_{ki}},+_{D_{ki}}\rangle+|-_{C_{ki}},-_{D_{ki}}\rangle\right)
\end{equation}
and
\begin{equation}
    \left|\phi^-_{C_{ki}D_{ki}}\right\rangle=\dfrac{1}{\sqrt{2}}\left(|+_{C_{ki}},-_{D_{ki}}\rangle+|-_{C_{ki}},+_{D_{ki}}\rangle\right).
\end{equation}
This means that if we perform $X$ basis measurements on all the qubits of the ancilla Bell states, we will obtain even parity results $\left|+_{C_{ki}} +_{D_{ki}}\right\rangle$
or $\left|-_{C_{ki}} -_{D_{ki}}\right\rangle$
for all the unflipped Bell pairs
and odd parity results
$\left|+_{C_{ki}} -_{D_{ki}}\right\rangle$ or $\left|-_{C_{ki}} +_{D_{ki}}\right\rangle$
for the Bell pairs flipped by the CZ operations.
This way, an $X$ basis measurement will tell us which memory qubits have the excitations. It will reveal the time bin in which the photon arrived, and also serve as a projective measurement for determining its spatial mode.
This will collapse the sum over the spatial modes in state \eqref{abcd}, giving us the resulting state of the memory qubits 
\begin{align}
\sigma_{\overline{AB},mq}(\textbf{h}) =\frac{1}{2}\sum_{s\in\{1,2\}}\left|\chi_{\overline{AB},mq}^{(s)}(\textbf{h})\right\rangle\left\langle\chi_{\overline{AB},mq}^{(s)}(\textbf{h})\right|
\end{align}
with corresponding probabilities (conditional upon there being an incoming photon) 
\begin{align}
p_{q}(\theta)=\eta^2_{q}(\theta).
\end{align}
Having found the spatial mode, the remaining task is to determine if the photon is in the symmetric or anti-symmetric combination of mode $q$ collected at the two sites.
Now, the memory qubits corresponding to the unflipped Bell states (i.e., the ones found to have even parity) are all in $\left|0\right\rangle$. These contain no useful information and we can drop them going forward. 
This leaves only the memory qubits corresponding to the Bell pairs found to be in $|\phi^-\rangle$ (i.e. the ones for which we obtained odd parity results).
Let $N_m\in\{1,...,\overline{M}\}$ be the number of such Bell pairs found to have flipped into $\left| \phi^- \right\rangle$ after the CZ gates step for a given time bin $m$ in which the photon arrived.
We can now define subsystems $E$ and $F$ containing only the $2N_m$ remaining memory qubits, and we adopt the labeling $\gamma\in\{E,F\}$ for these new systems. Then we construct the reduced state consisting of $2N_m$ entangled qubits, and this has the form
\begin{align}
\upsilon_{EF,mq}(\textbf{h}) = \frac{1}{2}\sum_{ s\in\{1,2\}} \left| \omega_{EF,mq}^{(s)}(\textbf{h})\right\rangle\left\langle \omega_{EF,mq}^{(s)}(\textbf{h})\right|, 
\end{align}
where
\begin{align}
    \left| \omega_{EF,mq}^{(s)}(\mathbf{h}) \right\rangle = \frac{1}{\sqrt 2}\left( e^{i\beta x_s}|\textbf{0}_{E_{mq}},\textbf{1}_{F_{mq}}\rangle 
    \right. \nonumber \\ \left.
    + f(\textbf{h}_{mq}) e^{-i\beta x_s}|\textbf{1}_{E_{mq}},\textbf{0}_{F_{mq}}\rangle\right),
    \label{memket}
\end{align}
with new kets defined as
\begin{align}
    \left|\textbf{0}_{E_{mq}},\textbf{1}_{F_{mq}}\right\rangle = \bigotimes_{\mu=1}^{N_m} |0_{E_{\mu q}},1_{F_{\mu q}}\rangle
\label{ef01-ket}
\end{align}
and
\begin{align}
    \left|\textbf{1}_{E_{mq}},\textbf{0}_{F_{mq}}\right\rangle = \bigotimes_{\mu=1}^{N_m} |1_{E_{\mu q}},0_{F_{\mu q}}\rangle.
\label{ef10-ket}
\end{align}
Note that we have defined the states \eqref{memket} in terms of $x_s$. We will make the substitutions $x_1=\theta$ and $x_2=-\theta$ for our 2-point-source separation-estimation problem later.

To determine if the incoming photon was in the symmetric or anti-symmetric combination of the $q$ mode signals collected at the two sites, we need to measure this in the basis
 \begin{align}
\left|\zeta^\pm_{EF,mq}\right\rangle =\frac{1}{\sqrt{2}} (|\textbf{0}_{E_{mq}},\textbf{1}_{F_{mq}}\rangle \pm |\textbf{1}_{E_{mq}},\textbf{0}_{F_{mq}}\rangle),
     \label{zetapm}
 \end{align}
The measurement interpretations will depend on the value of $f(\textbf{h}_{mq})$ which we know from the $X$ basis measurement outcomes of the photonic qubit measurements described in the encoding section.
Since the parity checks of the Bell pairs have by now revealed the temporal mode $m$ of the incoming photon and collapsed the state to a particular spatial mode $q$, we now revert to our stored results for our $X$ basis measurement of the corresponding photonic qubits at the end of the encoding section. To remind the reader, we obtain
$f(\textbf{h}_{mq}) =\pm 1$ for even/odd parity results, respectively.

Now, when $f(\textbf{h}_{mq}) =1$, measuring in the $|\zeta_\pm\rangle$ states corresponds to measuring in the basis $|\phi_{AB,q}^\pm\rangle$, stated in equations \eqref{phipm}. Whereas when $f(\textbf{h}_{mq}) =-1$, then these flip due to the minus sign, so the $|\zeta_{EF,mq}^\pm\rangle$ measurement corresponds to measuring in $|\phi_{AB,q}^\mp\rangle$.

To obtain the measurement probabilities, we write \eqref{memket} in terms of $\left|\zeta^\pm_{EF,mq}\right\rangle$, obtaining
\begin{subequations}
\begin{align}
    \left|\omega_{EF,mq} ^{(s)}(\textbf{h})\right\rangle &= \frac{1}{2}\left(e^{i\beta x_s}+ f(\textbf{h}_{mq}) e^{-i\beta x_s}\right)  \left|\zeta^+_{EF,mq}\right\rangle \nonumber \\ & \quad + \frac{1}{2}\left(e^{i\beta x_s}- f(\textbf{h}_{mq}) e^{-i\beta x_s}\right)  \left|\zeta^- _{EF,mq}\right\rangle  \nonumber \\
    &=  c_+^{(s)}(\mathbf{h}_{mq}) \left|\zeta^+_{EF,mq}\right\rangle + c_-^{(s)}(\mathbf{h}_{mq})   \left|\zeta^-_{EF,mq}\right\rangle,
    \label{omegamq}
\end{align} 
\end{subequations}
where 
\begin{align}
    c_\pm^{(s)}(\mathbf{h}_{mq}) \equiv \frac{1}{2}\left(e^{i\beta x_s} \pm f(\textbf{h}_{mq}) e^{-i\beta x_s}\right)
    \label{cpm}
\end{align}
are the probability amplitudes of obtaining the measurement outcome $|\zeta^+ _{EF,mq}\rangle$ or $|\zeta^- _{EF,mq}\rangle$, respectively.
 
At this point, the task of measuring \eqref{memket} in the \eqref{zetapm} states is, in principle, a GHZ state measurement. This can be done by measuring each qubit in the $X$-basis and then considering the parities of the outcomes associated with the results for each pair of qubits at both sites corresponding to the same digit of the binary encoding of the spatio-temporal mode $(m, q)$. 
We show in appendix \ref{GHZ-measurement} that an even number of odd parity results i.e. $|+_{E_{\mu q}},-_{F_{\mu q}}\rangle$ or $|-_{E_{\mu q}}, +_{F_{\mu q}}\rangle$
means finding the state to be $|\zeta^+ _{mq}\rangle$, 
and an odd number of odd parity results signifies obtaining $|\zeta^- _{EF,mq}\rangle$ as our measurement outcome.
Putting together the pieces along with the two cases for $f(\textbf{h}_{mq})$, we obtain the conditional probabilities $p_\pm(x_s)$ that given an incoming photon is in spatial mode $q$, it is in the symmetric or anti-symmetric combinations $|\phi_{AB,q}^\pm\rangle$ of the signals collected at the two sites
\begin{align}
p_+(x_s) &= [c_+^{(s)}(\textbf h)]^2 = \cos^2(\beta x_s) \\
p_-(x_s) &= [c_+^{(s)}(\textbf h)]^2 = \sin^2(\beta x_s)
\end{align}
for $f(\textbf{h}_{mq})=1$, and
\begin{align}
p_+(x_s) &= [c_-^{(s)}(\textbf h)]^2 = \cos^2(\beta x_s) \\
p_-(x_s) &= [c_-^{(s)}(\textbf h)]^2 = \sin^2(\beta x_s).
\end{align}
For $f(\textbf{h}_{mq})=-1$.

Overall, we obtain the full probabilities for a photon originating from a source at $x_s$ to be in the modes $|\phi_{AB,q}^\pm\rangle$:
\begin{align}
    P_{q+}(x_s) &= p_+(x_s) P_q(x_s) = \cos^2 (\beta x_s) \eta_q ^2(x_s) \nonumber \\
    P_{q-}(x_s) &= p_-(x_s) P_q(x_s) = \sin^2 (\beta x_s) \eta_q ^2(x_s).
\end{align}

The full probability of the symmetric and anti-symmetric states from both stars will be their brightness-weighted sum. Recalling that $x_1=\theta$ and $x_2=-\theta$, we obtain $P_{q\pm}(x_1) = P_{q\pm}(x_2) = P_{q\pm}(\theta)$ since $\eta_q ^2(\theta)\cos^2(\beta\theta)  = \eta_q ^2(-\theta) \cos^2(-\beta\theta) $ and
$\eta_q ^2(\theta)\sin^2(\beta\theta)  = \eta_q ^2(-\theta) \sin^2(-\beta\theta) $. The total probability for a photon from either of the two stars to be in modes \eqref{phipm} will therefore be
\begin{align}
P_{q\pm}(\theta) = \frac{1}{2}\sum_{s\in\{1,2\}} P_{q\pm}(x_s) = \frac{1}{2}[P_{q\pm}(\theta) + P_{q\pm}(-\theta)] 
\end{align} 
That is,
\begin{align}
    P_{q+}(\theta) = \eta_q ^2(\theta) \cos^2(\beta\theta) 
    \label{pq1}
\end{align}
and
\begin{align}
    P_{q-}(\theta) = \eta_q ^2(\theta) \sin^2(\beta\theta) ,
    \label{pq2}
\end{align}
which are exactly the probabilities \eqref{pplus1} and \eqref{pminus1}. These are exactly the probabilities from~\cite{Sajjad2023}, except of course for the change that we are obtaining $\eta_q(\theta)$ in place of $\Gamma_q(\theta)$ due to the fact that we are only collecting the first $K$ modes. As we mentioned earlier when we introduced $\eta_q(\theta)$ in \eqref{eta-definition}, $\eta_q(\theta) \to \Gamma_q(\theta)$ when $K$ is sufficiently large. We have thus reproduced the probabilities of the different measurement outcomes from~\cite{Sajjad2023} with our entanglement-based protocol as we set out to do.

\section{Quantum limit of precision for estimating two-source separation}\label{sec:quantumlimit_twosource}

The maximum amount of physically allowed information we can obtain about an unknown parameter is given by the quantum Fisher information (QFI)~\cite{Helstrom1976}. For the two-source separation problem, Tsang \textit{et al.} in~\cite{Tsang2016b} derive a general expression for the QFI matrix. For our convention of the stars positioned at $\pm\theta$, the component associated with the separation is 
\begin{align}
    \mathcal{K}=4N\int \text{d}x \left| \frac{\partial \psi(x)}{\partial x} \right|^2.
    \label{qfi2}
\end{align}
where $\psi(x)$ is the PSF, and $N$ is the total number of incoming photons during the integration time.
Taking the combined PSF \eqref{compound-psf-2-apertures-hard} for our system of 2 hard apertures, and setting $\beta=\pi r/\sigma$, where $r=2\beta/\delta$ is the ratio between the mid-point separation $2\beta$ (between the two apertures) and the individual aperture size $\delta =2\pi/\sigma$, gives the following QFI
    \begin{align}
    \mathcal{K} = \frac{4 \pi^2 N}{3\sigma^2}(3r^2+1).
    \end{align}
This result was originally obtained in~\cite{Sajjad2023}.
(Do note that in their conventions, the two apertures are at $\pm\beta/2$ instead of $\pm\beta$, hence their $r=\beta/\delta$, whereas we have $r=2\beta/\delta$ in order to have the physically same $r$ as them with the same QFI and CFI expressions.).

The performance of a particular measurement for determining  an unknown parameter is given by the classical Fisher information (CFI). For a single parameter problem, the CFI for a discrete random variable is given by
\begin{align}
    \mathcal{J}(\theta)= N\sum_i \frac{1}{P_i}\left(\frac{\partial P_i}{\partial \theta}\right)^2,
\end{align}
where $P_i$ is the probability of getting outcome $i$ and $N$ the number of copies of the state being measured.
In our problem, the CFI for outcome probabilities \eqref{pq1} and \eqref{pq2} is
 \begin{subequations}
        \begin{align}\label{cfiq}
        \mathcal{J}_q(\theta)&= \frac{N_K}{P_{q+}(\theta)}
        \left(\frac{\partial P_{q+}(\theta)}{\partial \theta}\right)^2 
       +\frac{N_K}{P_{q-}(\theta)}\left(\frac{\partial P_{q-}(\theta)}{\partial \theta}\right)^2 \\
        &= 4N_K \left[\beta^2 \eta_q ^2(\theta)
        +\left[\eta_q ^\prime(\theta)\right]^2\right],
        \end{align}
    \end{subequations}
where as per \eqref{eta-definition}, \small{$\eta_q(\theta) \equiv \Gamma_q(\theta)/\sqrt{\sum_{l=0}^{K-1} \Gamma_l ^2(\theta)}$}
and $\eta_q ^\prime(\theta) = \partial\eta_q(\theta)/\partial\theta$. $N_K$ is the (average) number of incoming photons in the first $K$ spatial modes and is given by \eqref{N_k-def}.

\begin{figure}
    \centering
    \includegraphics[scale=0.43]{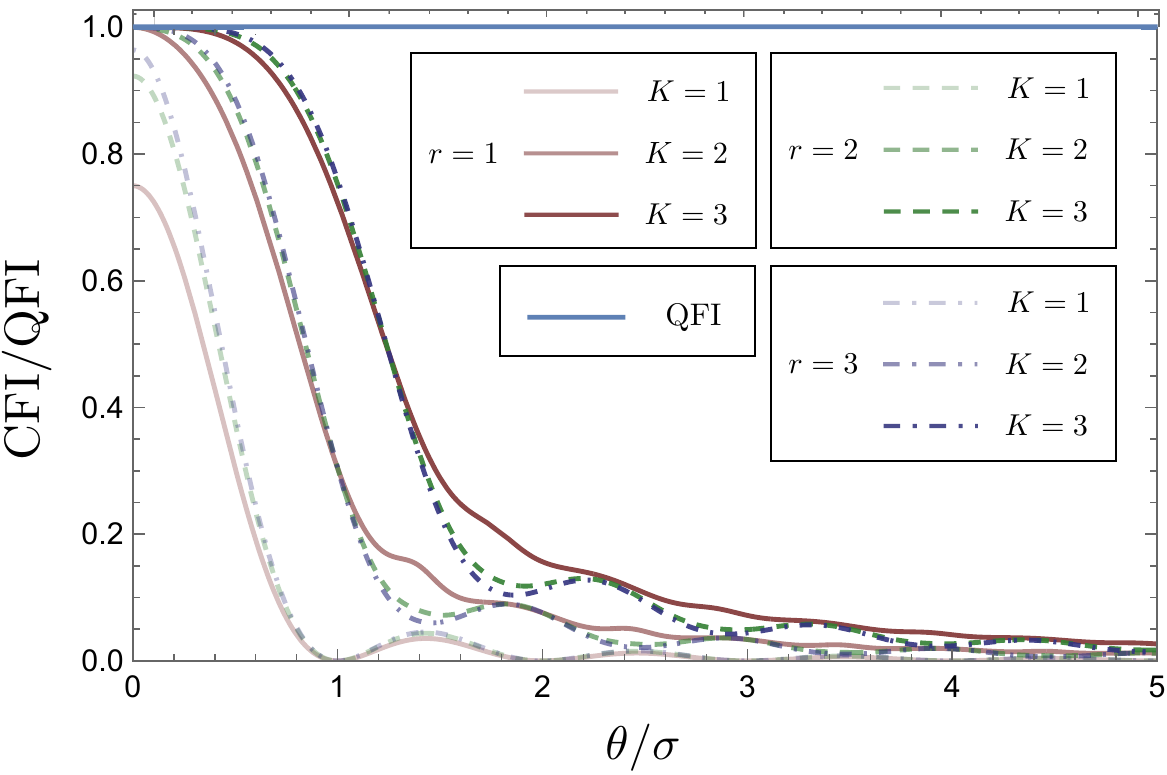}
    \caption{Plots of the normalized CFI with respect to the QFI for three different maximum numbers of spatial modes $K$ and increasing ratios: $r=1$ (solid lines), $r=2$ (dashed lines) and $r=3$ (dot-dashed lines).}\label{fig3}
\end{figure}

We can now add the CFI contributions over all the spatial modes
\small{
\begin{subequations}
\begin{align}
    \mathcal{J}_{K\text{-SPADE}}(\theta) &=\sum_{q=0}^{K-1}\mathcal{J}_q(\theta)  \\
    &= 4N_K \left[\frac{\pi^2 r^2}{\sigma^2}
       +\sum_{q=0}^{K-1} \left[\eta_q ^\prime(\theta)\right]^2\right] \\
               &= 4N \left\{ \left[\sum_{q=0}^{K-1} \frac{\pi^2 r^2}{\sigma^2} \Gamma_q ^2(\theta)
       +\left[\Gamma_q ^\prime(\theta)\right]^2\right] \right. \nonumber\\  &\quad\quad\quad\quad\quad\quad\left.-\frac{\left(\sum_{i=0}^{K-1} \Gamma_i(\theta) \Gamma_i ^\prime(\theta)\right)^2}{\sum_{j=0}^{K-1} \Gamma_j ^2(\theta)}\right\}, 
\end{align}    
\end{subequations}}

where in the second step, we have used $\sum_{q=0}^{K-1} \eta_q ^2(\theta) =1$ from the normalization condition and 
$\beta=\pi r/\sigma$. In the last step, we have used \eqref{N_k-def} and \eqref{eta-definition} to express the result in terms of the total incoming photon flux $N$ (in all the spatial modes) and the single-aperture correlation functions $\Gamma_q(\theta)$ and their derivatives.
In the large $K$ limit, $\sum_{q=0}^{K-1} \Gamma_q(\theta) \to 1$ from normalization. Then for our special case of a hard aperture with a sinc function PSF \eqref{PSF-sinc-def}, and taking the mode functions to be the normalized sinc-bessel modes $\phi_q(\theta)$, we find that when $K\to\infty$, the total CFI approaches the QFI~\cite{Sajjad2023}
\begin{align}
\lim_{K\to\infty} \mathcal{J}_{K\text{-SPADE}}(\theta)
=\frac{4 \pi^2 N}{3\sigma^2}(3r^2+1) =\mathcal{K},
    \end{align}
In Fig. \ref{fig3}, the CFI is plotted normalized with respect to the QFI for different values of $K$ and $r$. One can see how increasing these two parameters the CFI approaches the QFI in the small $\theta/\sigma$ limit.

Note in the QFI (and CFI for the $K\to\infty$ case)
that $4N\pi^2/3\sigma^2$ is in fact equal to the QFI for just a single aperture~\cite{Sajjad2023,kerviche2017,Paur:16}. So having two apertures is resulting in a gain by a factor of $3r^2+1$.
The $3r^2$ term thus contains the extra information arising from having two apertures and employing baseline interferometry, whereas the other piece is just the single-aperture QFI. In the above expressions, it is straightforward to see that resolving the light collected by the two apertures into spatial modes contributes to the latter piece whereas the former is obtained by combining the signals from the two telescopes~\cite{Sajjad2023}.

While we have derived the QFI and CFI result for the one-dimensional version of the problem, it turns out that this has a rather nice interpretation for the two-dimensional case. Let us assume that our two apertures are still located at positions $\pm\beta$ along the horizontal axis of the aperture plane, but each of these is now a square of length $\delta$ instead of being a one-dimensional rect function. The 2-dimensional equally bright two-point separation estimation problem has been worked out in~\cite{Ang2016}, where they calculate the QFI matrix in cartesian coordinates for the $x$ and $y$ components of the separation with a single aperture. Combining that with the multiple-aperture framework introduced in~\cite{Sajjad2023}, we can calculate the QFI for the $x$-component of the separation. We outline this calculation in appendix \ref{2-d-appendix}. Moreover, we can follow the same protocol described above in order to measure the incoming photons in the pairwise combinations of the (2-dimensional) local modes of the two apertures, and in the large $K$ limit, it will again attain the QFI which will be given by the same expression $4N\pi^2(3r^2+1)/\sigma^2$~\footnote{The proof that the pairwise measurement with 2-dimensional apertures attains the QFI is beyond the scope of this work and will be shown in a future publication.}.

Now, recall that $\sigma =2\pi/\delta$, with $\delta$ the size and $r=2\beta/\delta$.
Also, the total number $N$ of photons collected during the integration time will be proportional to the area of the apertures $\delta^2$.
So all in all, the $4N\pi^2/\sigma^2$ factor will scale as $\delta^4$.
The $3r^2$ term in the QFI will scale as $\delta^2\beta^2$, whereas the other term will be proportional to $\delta^4$.
This tells us that as $\delta$ gets larger (or equivalently as $r$ is small and closer to unity), we will get more information from the spatial mode resolution of the light within each aperture; whereas in the other extreme of $\delta$ being small ($r \gg \beta$), loading the photons contained within the local fundamental (PSF) spatial modes of each telescope's primary mirror into the quantum memory (i.e., no mode sorter, or $K=1$) will already achieve performance close to the QFI. A fully-general statement to this effect---regarding the relative-information-contributions from the local spatial modes versus the baseline---to more complex quantitative imaging problems such as multi-source localization, moment estimation, object classification, and change detection, etc., are left open for future work.

\section{Generalization to multiple telescopes and more complex imaging problems}\label{sec:generalproblem}

Since the QFI-attaining measurement for any single-parameter estimation task---no matter how complex the scene and its configuration---is a von Neumann projective measurement, given the one-photon-per-temporal-conherence-interval assumption, the QFI attaining measurement can always be expressed as a linear interferometric unitary transformation among all the locally-collected spatial modes across all the telescopes in the baseline, followed by shot-noise-limited photon detection on the output modes. We now discuss how the protocol we present in this paper can be generalized to measuring the incoming photons in such an arbitrary linear combination basis of the local modes collected by any number of telescopes. Say we have $n$ telescopes and we are collecting $K$ mutually-orthogonal spatial modes at each site. And our goal is to measure the photons in some arbitrary set of mutually orthogonal linear combinations of these $nK$ signals. If we are carrying out such a measurement by bringing all the $nK$ mode signals to a central location through single mode fibers, then we can use a beam splitter circuit to create an interferometer that mixes all the signals in the desired set of linear combinations. It has been shown in~\cite{Clements2016} that we can create such an interferometer that gives us any desired unitary transformation between the $nK$ inputs and the same number of outputs by employing $nK(nK-1)/2$ Mach Zehnder interferometers (MZIs). Since an MZI comprises of two 50-50 beam splitters and two tunable phase shifts, this means a total of $nK(nK-1)$ 50-50 beam splitters and the same number of phase shifts is needed.
If we are using pre-shared entanglement to accomplish this losslessly, instead of using (lossy) single mode fibers and beam splitters, then we need the following ingredients:
\begin{enumerate}
\item A general version of interpretting the parities of the $X$ basis measurement results of the photonic qubits and how they translate into relative signs between the various terms in the quantum state. In our two telescope example, where we are only mimicking the classical interferometer that mixes the signals from the two sites associated with the same spatial mode, we only have to consider the parity of the $X$ basis results for a single spatial mode at both sites. In the most general case where we also mix multiple spatial modes together, we will have to consider the $X$ basis results of several spatial modes at various sites simultaneously. Therefore, instead of just a single sign $f(\textbf{h}_{mq})$, we will have to keep track of several such signs in a slightly elaborate (but easily doable) book keeping exercise.
\item $n$-site and $K$-spatial mode generalization of our heralding procedure in the decoding section for determining the spatio-temporal mode of an incoming photon with ancilla Bell states, once the state of the incoming photons has been loaded on to quantum memories. It is straightforward to see that using $CZ$ gates in the same way but employing a GHZ state instead of a Bell pair can easily accomplish this. We will shortly describe this more explicitly.
\item An operation on a pair of logical qubits comprising several atomic memories, that mimics the action of a fifty-fifty beam splitter between two optical modes whose quantum state resides in the span of one photon spread across the two modes. These logical qubits may be at the same location or at two distant sites, in which case accomplishing this requires an entangled state shared across the two locations. We will shortly make this statement more precise and show that we can indeed accomplish this.
 \item The ability to apply any arbitrary phase shift to an optical mode, or a single-qubit phase after the light-to-qubit transduction explained in the paper. This can be accomplished easily in either domain.
\item (Optional) An $n$ site and multiple spatial mode generalization of our procedure of using Bell states ancillas to realize the nonlocal version of the Clements {\em et al.} circuit~\cite{Clements2016} more resource-efficiently by employing multi-partite entangled resources instead of only using pairwise Bell states and mimicking fifty-fifty beam splitters.
\end{enumerate}
We now clarify the generalization of the heralding procedure with GHZ states. Our heralding procedure with Bell states in the decoding section is based on the fact that if we have a Bell state $\frac{1}{\sqrt{2}}[|00\ra +|1 1\ra]$,
then applying a $Z$ gate on one of the two qubits imparts a minus sign on the second term, flipping the Bell state to $\frac{1}{\sqrt{2}}[|00\ra -|1 1\ra]$.
Therefore, if we have a pair of memory qubits in the $|1 0\ra$ or $|0 1\ra$, then applying a $CZ$ gate from one of these memories to one qubit in the Bell pair, and another $CZ$ from the other memory qubit to the second qubit in the Bell pair, flips the sign of the Bell state. But if we have $|00\ra$ in the two memories, the Bell state will remain unflipped.

This logic clearly also holds when we have a GHZ state. Consider the states
\be
|GHZ^\pm _n\ra \equiv \frac{1}{\sqrt{2}}[|0_{1^\prime} 0_{2^\prime} \ldots 0_{n ^\prime}\ra \pm |1_{1^\prime} 1_{2^\prime}\ldots 1_{n^\prime}\ra].
\ee
Then
\begin{eqnarray}
V_{1 1^\prime} |1_1 0_2 0_3\ldots 0_n\ra |GHZ^+ _n\ra
&=& V_{2 2^\prime} |1_1 0_2 0_3\ldots 0_n\ra |GHZ^+ _n\ra \nn \\
&=&\ldots \nn \\
&=& V_{n n^\prime} |0_1 0_2\ldots 0_{n-1} 1_n\ra |GHZ^+ _n\ra \nn \\
&=& |GHZ^- _n\ra,
\end{eqnarray}
where $V_{j, j^\prime}$ is a $CZ$ between qubits $j$ and $j^\prime$.
We can thus generalize our heralding procedure by employing GHZ states.

 Also, in our two-telescope example, we determine the temporal and spatial mode simultaneously since we are only measuring in linear combinations involving the same spatial mode collected at both sites. In a general receiver where we are measuring in a basis that mixes two or more spatial modes, we may have a heralding procedure that only determines the temporal mode through the heralding step, while leaving the spatial mode completely undetermined, or only narrows down to certain combinations of the spatial modes.
For example, say we have 4 telescopes with $K=4$. If for each $q$, we wish to measure the incoming photon from the scene in some linear combination of the mode $q$ state associated with each site, then we are not mixing the different spatial modes with each other. In that case, a heralding procedure with a 4-component GHZ state for each set of memory qubits (corresponding to  a given digit in the binary representation of the temporal modes) can be used to simultaneously determine the temporal and spatial mode just as in our pairwise receiver for two telescopes.
However, if we wish to measure the incoming photons in some combination of the different spatial modes, we will not want to employ a heralding procedure that collapses tthe state on to a single spatial mode number $q$. In such a situation, we will design the heralding procedure according to our measurement basis. If there is no simplification where the heralding procedure can help us narrow down to a subspace spanned by a subset of the states in the measurement basis, the most general, brute-force approach would involve a $Kn$-component GHZ state that only determines the temporal mode while keeping the entire spatial mode related structure of the quantum state in tact.

We now show how to mimic a fifty-fifty beam splitter on the Hilbert space spanned by one photon across two modes, $\left\{ |1_A 0_B\rangle, |0_A 1_B\rangle\right\}$. Consider the input modes $A$ and $B$ of a 50-50 beam splitter with output modes $A'$ and $B'$. The beam splitter will have the following action:
\begin{align}
a |1_A 0_B\rangle & + b |0_A 1_B\rangle \\
    & \downarrow \nonumber \\ 
    \frac{a+b}{\sqrt{2}} |1_{A'} 0_{B'}\rangle 
    &+ \frac{a-b}{\sqrt{2}}|0_{A'} 1_{B'}\rangle.
\label{beam-splitter-action}
\end{align}
In our entanglement-based receiver, we load the photonic state on to quantum memories. With the logarithmic compression of Khabiboulline {\em et al.}, a (single-rail) photonic qubit for each spatial mode at each site associated with temporal mode $m$ is loaded on to $n_m$ (i.e. the number of $1$s in the binary code of $m$ at the same location. If there is no incoming photon in the said spatial mode at that site in temporal mode $m$, then these $n_m$ memory qubits remain in $|0\ra$, and if there is a photon, then all of them are flipped to $|1\ra$ in the loading process. These $n_m$ memory qubits therefore effectively serve as logical qubits associated with the corresponding spatial mode.
To obtain the quantum version of the Clements {\em et al.} circuit~\cite{Clements2016}, we need to apply
an operation that gives the operation (\ref{beam-splitter-action}) on two such logical qubits, wich may be at the same location or two distant sites.
If the states of the input optical modes $A$ and $B$ are transfered on to logical qubits $\overline{E}$ and $\overline{F}$,
then the above (beam splitter equivalent) operation can be accomplished by carrying out the following sequence
\begin{enumerate}
\item A CNOT gate from logical qubit $\overline{E}$ to $\overline{F}$.
\item The Hadamard gate on logical qubit $\overline{E}$.
\item A CNOT gate from logical qubit $\overline{E}$ to $\overline{F}$.
\end{enumerate}
When the two qubits are at two different sites, the CNOT operations need to be carried out through gate teleportation, which consumes one pre-shared two-qubit Bell state, each, in a way identical to that in Ref. ~\cite{gottesman1999quantum}.
We have thus shown how to mimic a 50-50 beam splitter across two sites.
We can therefore use this and apply the framework in~\cite{Clements2016} to mimic any nonlocal multimode beam splitter circuit to measure the incoming collected optical field across $n$ remotely-located modes in any arbitrary collective basis. This is true as long as the quantum state of those modes lives in the unary subspace, i.e., one photon spread across the $n$ modes. This condition is satisfied as part of the premise of the problem setup as described above.

\section{Conclusion}\label{sec:conclusions}

Shared entanglement offers a very interesting and potentially promising direction for long-baseline interferometry over large distances that are currently not feasible with traditional techniques due to losses and other practical constraints.
However, this still remains a largely unexplored direction with a few pioneering works such as~\cite{Gottesman2012} and~\cite{Khabiboulline2018}
proposing how to estimate the phase difference between two sites using shared entanglement, which amounts to estimating the location of a single monochromatic point source.
The former work introduces this concept and the latter provides a way to do this with substantially fewer entanglement resources with a logarithmic compression of the information-bearing light into atomic qubit registers.
Here we have taken a step towards combining these ideas with a suite of recent research on pre-detection spatial mode sorting to achieve quantum-limited spatial resolution of a single telescope, to addressing general quantitative imaging and parameter estimation tasks with long-baseline telescope arrays. Our approach involves sorting the light collected at each telescope site into a few highest-order local spatial modes, storing the information bearing light in each sorted mode into atomic qubit memory registers, and using shared entanglement among the telescope sites to effect the multimode linear optical mode mixing prior to detection, but without bringing the light from the telescopes to one common location. For the problem of estimating the separation between two point sources, leveraging the result from ~\cite{Sajjad2023}---which proved that the QFI-attaining measurement involves measuring the incoming light in terms of the sums and differences of the modal signal amplitudes received at the two telescopes, we fully work out our qubit-efficient interferometric technique to realize this QFI-attaining measurement in a nonlocal fashion. We also outline how our protocol can be generalized to an entanglement based recipe to attain the quantum limit for any quantitative imaging or parameter estimation problem. We hope this will inspire more research towards eventually building a real long baseline interferometric system that employs high-fidelity shared entanglement, which brings together spatial-mode sorting, programmable linear optics, efficient light-matter interfaces to spin qubits, and high-fidelity quantum logic among spin qubits in one working system.

\section{Acknowledgements}
The authors thank Michael Grace, Gabe Richardson, Brittany McClinton, Jayadev Rajagopal, Ryan Lau, and Stephen Ridgway for helpful discussions, and Prajit Dhara for comments on the manuscript. This work was co-funded by AFOSR grant number FA9550-22-1-0180 and NASA grant number 80NSSC22K1030.

\onecolumngrid

\appendix

\section{Derivation of the $M$-copy state of the incoming light state}
\label{A1}
Here we present a more detailed derivation of Eq. \eqref{rhoabM} from the main text. The key to understand these steps is that expand the state time bin at a time, so that in every step we get rid of all the non-linear terms of the mean photon number $\epsilon$. For example, we can begin by computing the tensor product of two copies of $\rho_{AB}$:
\begin{subequations}
    \begin{align}
    \rho^{\otimes 2}_{AB} &= \left[(1-\epsilon)\rho_{0,{AB}}+\epsilon\rho_{1,AB}\right]^{\otimes 2} \\
    &= (1-\epsilon)^2\rho_{0,{AB}}^{\otimes 2} + (1-\epsilon)\epsilon(\rho_{0,AB}\otimes\rho_{1,AB}+\rho_{1,AB}\otimes\rho_{0,AB}) +\epsilon^2 \rho_{1,{AB}}^{\otimes 2} \\
    &= (1-2\epsilon)\rho_{0,AB}^{\otimes 2}+\epsilon(\rho_{0,AB}\otimes\rho_{1,AB}+\rho_{1,AB}\otimes\rho_{0,AB})+\epsilon^2(\rho_{0,{AB}}^{\otimes 2}-\rho_{0,AB}\otimes\rho_{1,AB}-\rho_{1,AB}\otimes\rho_{0,AB}+\rho_{1,{AB}}^{\otimes 2}).
\end{align}
\end{subequations}
Since the non-linear therms of $\epsilon$ are neglected, we can drop the $\epsilon^2$ term. Thus,
\begin{align}
    \rho^{\otimes 2}_{AB}
    &\approx (1-2\epsilon)\rho_{0,AB}^{\otimes 2}+\epsilon(\rho_{0,AB}\otimes\rho_{1,AB}+\rho_{1,AB}\otimes\rho_{0,AB}).
\end{align}
Now we repeat this procedure over the $M$ time bins:
\begin{subequations}
\begin{align}
    \rho^{\otimes M}_{AB} &= \left[(1-\epsilon)\rho_{0,{AB}}+\epsilon\rho_{1,AB}\right]^{\otimes M} \\ 
    &\approx\left[(1-2\epsilon)\rho_{0,AB}^{\otimes 2}+\epsilon(\rho_{0,AB}\otimes\rho_{1,AB}+\rho_{1,AB}\otimes\rho_{0,AB})\right]\otimes\rho^{\otimes (M-2)}\\ 
    &\approx\left[(1-3\epsilon)\rho_{0,AB}^{\otimes 3}+\epsilon(\rho_{1,AB}\otimes\rho_{0,AB}\otimes\rho_{0,AB}+\rho_{0,AB}\otimes\rho_{1,AB}\otimes\rho_{0,AB}+\rho_{0,AB}\otimes\rho_{0,AB}\otimes\rho_{1,AB})\right]\otimes\rho_{AB}^{\otimes (M-3)} \\
    & \vdots \nonumber \\ 
    &\approx (1-M\epsilon)\rho_{0,AB}^{\otimes M} + \epsilon\sum_{m=1}^M \rho_{1,AB,m} ,
\end{align}  
with 
\begin{align}
    \rho_{1,AB,m}\equiv\rho_{0,AB}^{\otimes (m-1)}\otimes\rho_{1,AB}\otimes\rho_{0,AB}^{\otimes (M-m)}
\end{align}
being defined to represent a one-photon state in the $m$-th time bin.
\end{subequations}

\section{Measurement in the $|\zeta^-_{EF,mq}\rangle$ states}
\label{GHZ-measurement}

Here we show how we can measure the state \eqref{memket} in the \eqref{zetapm} basis. At each site, we have $n_m$ qubits labeled by an index $\mu=1\ldots n_m$. We measure each qubit in the $X$ basis and for each $\mu$, compare whether the results at both sites are the same (i.e. $++$ or $--$ which we call even parity outcomes) or opposite (i.e. $+-$ or $-+$ which we call odd parity results). We will show that an even number of odd parity outcomes means we have $|\zeta^+ _{mq}\rangle$, 
and obtaining an odd number of odd parities means $|\zeta^- _{mq}\rangle$. 
Now, the state for a single qubit in the first term is
\begin{align}
    \left|0_{E_{\mu q}},1_{F_{\mu q}}\right\rangle 
    &= \frac{1}{2}\left[\left(\left|+_{E_{\mu q}}, +_{F_{\mu q}}\right\rangle -\left|-_{E_{\mu q}}, -_{F_{\mu q}}\right\rangle\right) \nonumber \right. \\ 
    &  \left. \quad + \left(\left|+_{E_{\mu q}}, -_{F_{\mu q}}\right\rangle -\left|-_{E_{\mu q}}, +_{F_{\mu q}}\right\rangle\right)\right] \\
    &= \frac{1}{\sqrt{2}} \left( \left|e_{EF,\mu q}\right\rangle +\left|o_{EF,\mu q}\right\rangle\right) 
\end{align} 
and in the second term, we have
 \begin{align}
     |1_{E_{\mu q}},0_{F_{\mu q}}\rangle 
    &= \frac{1}{2}\left[\left(\left|+_{E_{\mu q}}, +_{F_{\mu q}}\right\rangle -\left|-_{E_{\mu q}}, -_{F_{\mu q}}\right\rangle\right) \nonumber \right. \\ 
    &  \left. \quad - \left(\left|+_{E_{\mu q}}, -_{F_{\mu q}}\right\rangle -\left|-_{E_{\mu q}}, +_{F_{\mu q}}\right\rangle\right)\right] \\
    &= \frac{1}{\sqrt{2}} \left( \left|e_{EF,\mu q}\right\rangle - \left|o_{EF,\mu q}\right\rangle\right)
 \end{align}
where
 \begin{align}
      \left|e_{EF,\mu q}\right\rangle = \frac{1}{\sqrt{2}}(|+_{E_{\mu q}}, +_{F_{\mu q}}\rangle - |-_{E_{\mu q}}, -_{F_{\mu q}}\rangle)
      \label{Yeven}
 \end{align}
are the even parity states, and
\begin{align}
   \left|o_{EF,\mu q}\right\rangle = \frac{1}{\sqrt{2}}(|-_{E_{\mu q}}, +_{F_{\mu q}}\rangle - |+_{E_{\mu q}}, -_{F_{\mu q}}\rangle).
   \label{Yodd}
\end{align}
are the odd parity states. So the even parity state \eqref{Yeven} has the same sign in both $\left|0_{E_{\mu q}}, 1_{F_{\mu q}}\right\rangle$ and $\left|1_{E_{\mu q}}, 0_{F_{\mu q}}\right\rangle$, but the odd parity state \eqref{Yodd} has opposite signs.
 
Now we need to expand $N_m$ copies of $\left|0_{E_{\mu q}},1_{F_{\mu q}}\right\rangle$, and $N_m$ copies of $\left|1_{E_{\mu q}},0_{F_{\mu q}}\right\rangle$. In the former, i.e., $\left|0_{E_{\mu q}}, 1_{F_{\mu q}}\right\rangle^{\bigotimes N_m}$, all the terms will have positive signs since a single copy has positive signs in front of both $\left|e_{EF,\mu q}\right\rangle$ and $\left|o_{EF,\mu q}\right\rangle$. But in $\left|1_{E_{\mu q}}, 0_{F_{\mu q}}\right\rangle^{\bigotimes N_m}$, all terms with an odd number of $\left|o_{EF,\mu q}\right\rangle$ will have minus signs.
In other words, terms with an even number of $\left|o_{EF,\mu q}\right\rangle$ will have positive signs and will contribute to $|\zeta^+ _{mq}\rangle$, but terms having an odd number of $\left|o_{EF,\mu q}\right\rangle$ will have minus signs and will contribute to $|\zeta^- _{EF,mq}\rangle$. We have thus shown that we can measure whether we have $|\zeta^+ _{EF,mq}\rangle$ or $|\zeta^- _{EF,mq}\rangle$ depending on whether we have an even or odd number of odd parity results, respectively.

\section{The 2-dimensional QFI}
\label{2-d-appendix}

The two-dimensional version of the equally bright two-point separation estimation problem has been worked out in~\cite{Ang2016}, where they calculate the QFI matrix for the $x$ and $y$ components of the centroid and separation. They find that if the (2-dimensional) PSF has reflection symmetry in both $x$ and $y$ directions i.e. $\psi_{2d}(x, y) = \psi_{2d}(-x, y) = \psi_{2d}(x, -y) = \psi_{2d}(-x, -y)$, then the off-diagonal entries of the QFI matrix vanish and we can estimate the $x$ and $y$ components of the separation independently.
The QFI components are then given by
\begin{align}
\mathcal{K}_x &= 4N \int \text{d}x \text{d}y \left(\partial_x \psi_{2d}(x, y)\right)^2 \nonumber \\
\mathcal{K}_y &= 4N \int \text{d}x \text{d}y \left(\partial_y \psi_{2d}(x, y)\right)^2
\label{2d-QFI}
\end{align}
For simplicity, if we assume a square shaped aperture of length $\delta$, then our 2d PSF for a single aperture will simply be $\psi_{\text{2d, 1ap}}(x, y) = \psi(x) \psi(y)$,
where $\psi(x)$ is the 1-dimensional single-aperture PSF and was given in \eqref{PSF-sinc-def}. Now, if we consider two such apertures along the horizontal axis of our aperture plane at positions $\beta$ and $-\beta$, then following \eqref{psi-2ap-def}, the combined PSF will be
\begin{align}
\psi_{\text{2d, 2-ap}}(x) &= \sqrt{2} \cos(\beta x) \psi_{\text{2d, 1ap}}(x, y)
\label{psi-2ap-def-2d}
\end{align}
Inserting this in \eqref{2d-QFI}, we obtain $4N\pi^2(3r^2+1)/\sigma^2$ for the QFI for the horizontal component of the separation, same as the 1-dimensional QFI.

\twocolumngrid

%



\begin{thebibliography}{56}%
\makeatletter
\providecommand \@ifxundefined [1]{%
 \@ifx{#1\undefined}
}%
\providecommand \@ifnum [1]{%
 \ifnum #1\expandafter \@firstoftwo
 \else \expandafter \@secondoftwo
 \fi
}%
\providecommand \@ifx [1]{%
 \ifx #1\expandafter \@firstoftwo
 \else \expandafter \@secondoftwo
 \fi
}%
\providecommand \natexlab [1]{#1}%
\providecommand \enquote  [1]{``#1''}%
\providecommand \bibnamefont  [1]{#1}%
\providecommand \bibfnamefont [1]{#1}%
\providecommand \citenamefont [1]{#1}%
\providecommand \href@noop [0]{\@secondoftwo}%
\providecommand \href [0]{\begingroup \@sanitize@url \@href}%
\providecommand \@href[1]{\@@startlink{#1}\@@href}%
\providecommand \@@href[1]{\endgroup#1\@@endlink}%
\providecommand \@sanitize@url [0]{\catcode `\\12\catcode `\$12\catcode
  `\&12\catcode `\#12\catcode `\^12\catcode `\_12\catcode `\%12\relax}%
\providecommand \@@startlink[1]{}%
\providecommand \@@endlink[0]{}%
\providecommand \url  [0]{\begingroup\@sanitize@url \@url }%
\providecommand \@url [1]{\endgroup\@href {#1}{\urlprefix }}%
\providecommand \urlprefix  [0]{URL }%
\providecommand \Eprint [0]{\href }%
\providecommand \doibase [0]{http://dx.doi.org/}%
\providecommand \selectlanguage [0]{\@gobble}%
\providecommand \bibinfo  [0]{\@secondoftwo}%
\providecommand \bibfield  [0]{\@secondoftwo}%
\providecommand \translation [1]{[#1]}%
\providecommand \BibitemOpen [0]{}%
\providecommand \bibitemStop [0]{}%
\providecommand \bibitemNoStop [0]{.\EOS\space}%
\providecommand \EOS [0]{\spacefactor3000\relax}%
\providecommand \BibitemShut  [1]{\csname bibitem#1\endcsname}%
\let\auto@bib@innerbib\@empty
\bibitem [{\citenamefont {Lord~Rayleigh}(1879)}]{LordRayleigh1879}%
  \BibitemOpen
  \bibfield  {author} {\bibinfo {author} {\bibfnamefont {F.~R.~S.}\
  \bibnamefont {Lord~Rayleigh}},\ }\href {\doibase 10.1080/14786447908639684}
  {\bibfield  {journal} {\bibinfo  {journal} {Philosophical Magazine Series 5}\
  }\textbf {\bibinfo {volume} {8}},\ \bibinfo {pages} {261} (\bibinfo {year}
  {1879})}\BibitemShut {NoStop}%
\bibitem [{\citenamefont {Tsang}\ \emph
  {et~al.}(2016{\natexlab{a}})\citenamefont {Tsang}, \citenamefont {Nair},\
  and\ \citenamefont {Lu}}]{Tsang2016b}%
  \BibitemOpen
  \bibfield  {author} {\bibinfo {author} {\bibfnamefont {M.}~\bibnamefont
  {Tsang}}, \bibinfo {author} {\bibfnamefont {R.}~\bibnamefont {Nair}}, \ and\
  \bibinfo {author} {\bibfnamefont {X.-M.}\ \bibnamefont {Lu}},\ }\href
  {\doibase 10.1103/PhysRevX.6.031033} {\bibfield  {journal} {\bibinfo
  {journal} {Phys. Rev. X}\ }\textbf {\bibinfo {volume} {6}},\ \bibinfo {pages}
  {031033} (\bibinfo {year} {2016}{\natexlab{a}})}\BibitemShut {NoStop}%
\bibitem [{\citenamefont {Kerviche}\ \emph {et~al.}(2017)\citenamefont
  {Kerviche}, \citenamefont {Guha},\ and\ \citenamefont
  {Ashok}}]{kerviche2017}%
  \BibitemOpen
  \bibfield  {author} {\bibinfo {author} {\bibfnamefont {R.}~\bibnamefont
  {Kerviche}}, \bibinfo {author} {\bibfnamefont {S.}~\bibnamefont {Guha}}, \
  and\ \bibinfo {author} {\bibfnamefont {A.}~\bibnamefont {Ashok}},\ }in\ \href
  {\doibase 10.1109/ISIT.2017.8006566} {\emph {\bibinfo {booktitle} {2017 IEEE
  International Symposium on Information Theory (ISIT)}}}\ (\bibinfo {year}
  {2017})\ pp.\ \bibinfo {pages} {441--445}\BibitemShut {NoStop}%
\bibitem [{\citenamefont {{\u R}eh{\'a}{\u c}ek}\ \emph {et~al.}()\citenamefont
  {{\u R}eh{\'a}{\u c}ek}, \citenamefont {Pa{\'u}r}, \citenamefont {Stoklasa},
  \citenamefont {Motka}, \citenamefont {Hradil},\ and\ \citenamefont
  {S{\'a}nchez-Soto}}]{Rehacek2016}%
  \BibitemOpen
  \bibfield  {author} {\bibinfo {author} {\bibfnamefont {J.}~\bibnamefont {{\u
  R}eh{\'a}{\u c}ek}}, \bibinfo {author} {\bibfnamefont {M.}~\bibnamefont
  {Pa{\'u}r}}, \bibinfo {author} {\bibfnamefont {B.}~\bibnamefont {Stoklasa}},
  \bibinfo {author} {\bibfnamefont {L.}~\bibnamefont {Motka}}, \bibinfo
  {author} {\bibfnamefont {Z.}~\bibnamefont {Hradil}}, \ and\ \bibinfo {author}
  {\bibfnamefont {L.~L.}\ \bibnamefont {S{\'a}nchez-Soto}},\ }\href@noop {} {\
  }\Eprint {http://arxiv.org/abs/1607.05837} {arXiv:1607.05837 [quant-th]}
  \BibitemShut {NoStop}%
\bibitem [{\citenamefont {Ang}\ \emph {et~al.}(2017)\citenamefont {Ang},
  \citenamefont {Nair},\ and\ \citenamefont {Tsang}}]{Ang2016}%
  \BibitemOpen
  \bibfield  {author} {\bibinfo {author} {\bibfnamefont {S.~Z.}\ \bibnamefont
  {Ang}}, \bibinfo {author} {\bibfnamefont {R.}~\bibnamefont {Nair}}, \ and\
  \bibinfo {author} {\bibfnamefont {M.}~\bibnamefont {Tsang}},\ }\href
  {\doibase 10.1103/PhysRevA.95.063847} {\bibfield  {journal} {\bibinfo
  {journal} {Phys. Rev. A}\ }\textbf {\bibinfo {volume} {95}},\ \bibinfo
  {pages} {063847} (\bibinfo {year} {2017})}\BibitemShut {NoStop}%
\bibitem [{\citenamefont {Tsang}(2017)}]{tsang2017}%
  \BibitemOpen
  \bibfield  {author} {\bibinfo {author} {\bibfnamefont {M.}~\bibnamefont
  {Tsang}},\ }\href {\doibase 10.1088/1367-2630/aa60ee} {\bibfield  {journal}
  {\bibinfo  {journal} {New Journal of Physics}\ }\textbf {\bibinfo {volume}
  {19}},\ \bibinfo {pages} {023054} (\bibinfo {year} {2017})}\BibitemShut
  {NoStop}%
\bibitem [{\citenamefont {Tsang}(2019{\natexlab{a}})}]{Tsang2019}%
  \BibitemOpen
  \bibfield  {author} {\bibinfo {author} {\bibfnamefont {M.}~\bibnamefont
  {Tsang}},\ }\href@noop {} {\  (\bibinfo {year} {2019}{\natexlab{a}})},\
  \Eprint {http://arxiv.org/abs/arXiv:1906.02064} {arXiv:1906.02064}
  \BibitemShut {NoStop}%
\bibitem [{\citenamefont {Yu}\ and\ \citenamefont {Prasad}(2018)}]{Prasad2018}%
  \BibitemOpen
  \bibfield  {author} {\bibinfo {author} {\bibfnamefont {Z.}~\bibnamefont
  {Yu}}\ and\ \bibinfo {author} {\bibfnamefont {S.}~\bibnamefont {Prasad}},\
  }\href {\doibase 10.1103/PhysRevLett.121.180504} {\bibfield  {journal}
  {\bibinfo  {journal} {Phys. Rev. Lett.}\ }\textbf {\bibinfo {volume} {121}},\
  \bibinfo {pages} {180504} (\bibinfo {year} {2018})}\BibitemShut {NoStop}%
\bibitem [{\citenamefont {Prasad}\ and\ \citenamefont {Yu}(2019)}]{Prasad2019}%
  \BibitemOpen
  \bibfield  {author} {\bibinfo {author} {\bibfnamefont {S.}~\bibnamefont
  {Prasad}}\ and\ \bibinfo {author} {\bibfnamefont {Z.}~\bibnamefont {Yu}},\
  }\href {\doibase 10.1103/PhysRevA.99.022116} {\bibfield  {journal} {\bibinfo
  {journal} {Phys. Rev. A}\ }\textbf {\bibinfo {volume} {99}},\ \bibinfo
  {pages} {022116} (\bibinfo {year} {2019})}\BibitemShut {NoStop}%
\bibitem [{\citenamefont {{\v R}eha{\v c}ek}\ \emph {et~al.}(2017)\citenamefont
  {{\v R}eha{\v c}ek}, \citenamefont {Hradil}, \citenamefont {Stoklasa},
  \citenamefont {Pa{\'u}r}, \citenamefont {Grover}, \citenamefont {Krzic},\
  and\ \citenamefont {S{\'a}nchez-Soto}}]{Rehacek2017b}%
  \BibitemOpen
  \bibfield  {author} {\bibinfo {author} {\bibfnamefont {J.}~\bibnamefont {{\v
  R}eha{\v c}ek}}, \bibinfo {author} {\bibfnamefont {Z.}~\bibnamefont
  {Hradil}}, \bibinfo {author} {\bibfnamefont {B.}~\bibnamefont {Stoklasa}},
  \bibinfo {author} {\bibfnamefont {M.}~\bibnamefont {Pa{\'u}r}}, \bibinfo
  {author} {\bibfnamefont {J.}~\bibnamefont {Grover}}, \bibinfo {author}
  {\bibfnamefont {A.}~\bibnamefont {Krzic}}, \ and\ \bibinfo {author}
  {\bibfnamefont {L.~L.}\ \bibnamefont {S{\'a}nchez-Soto}},\ }\href@noop {}
  {\bibfield  {journal} {\bibinfo  {journal} {Phys. Rev. A}\ }\textbf {\bibinfo
  {volume} {96}},\ \bibinfo {pages} {062107} (\bibinfo {year}
  {2017})}\BibitemShut {NoStop}%
\bibitem [{\citenamefont {{\u R}eh{\'a}{\u c}ek}\ \emph
  {et~al.}(2018)\citenamefont {{\u R}eh{\'a}{\u c}ek}, \citenamefont {Hradil},
  \citenamefont {Koutn\'y}, \citenamefont {Grover}, \citenamefont {Krzic},\
  and\ \citenamefont {S\'anchez-Soto}}]{rehacek2018}%
  \BibitemOpen
  \bibfield  {author} {\bibinfo {author} {\bibfnamefont {J.}~\bibnamefont {{\u
  R}eh{\'a}{\u c}ek}}, \bibinfo {author} {\bibfnamefont {Z.}~\bibnamefont
  {Hradil}}, \bibinfo {author} {\bibfnamefont {D.}~\bibnamefont {Koutn\'y}},
  \bibinfo {author} {\bibfnamefont {J.}~\bibnamefont {Grover}}, \bibinfo
  {author} {\bibfnamefont {A.}~\bibnamefont {Krzic}}, \ and\ \bibinfo {author}
  {\bibfnamefont {L.~L.}\ \bibnamefont {S\'anchez-Soto}},\ }\href {\doibase
  10.1103/PhysRevA.98.012103} {\bibfield  {journal} {\bibinfo  {journal} {Phys.
  Rev. A}\ }\textbf {\bibinfo {volume} {98}},\ \bibinfo {pages} {012103}
  (\bibinfo {year} {2018})}\BibitemShut {NoStop}%
\bibitem [{\citenamefont {Prasad}(2020{\natexlab{a}})}]{Prasad_2020}%
  \BibitemOpen
  \bibfield  {author} {\bibinfo {author} {\bibfnamefont {S.}~\bibnamefont
  {Prasad}},\ }\href {\doibase 10.1088/1402-4896/ab573d} {\bibfield  {journal}
  {\bibinfo  {journal} {Physica Scripta}\ }\textbf {\bibinfo {volume} {95}},\
  \bibinfo {pages} {054004} (\bibinfo {year} {2020}{\natexlab{a}})}\BibitemShut
  {NoStop}%
\bibitem [{\citenamefont {Bisketzi}\ \emph {et~al.}(2019)\citenamefont
  {Bisketzi}, \citenamefont {Branford},\ and\ \citenamefont
  {Datta}}]{Bisketzi2019}%
  \BibitemOpen
  \bibfield  {author} {\bibinfo {author} {\bibfnamefont {E.}~\bibnamefont
  {Bisketzi}}, \bibinfo {author} {\bibfnamefont {D.}~\bibnamefont {Branford}},
  \ and\ \bibinfo {author} {\bibfnamefont {A.}~\bibnamefont {Datta}},\ }\href
  {\doibase 10.1088/1367-2630/ab58a0} {\bibfield  {journal} {\bibinfo
  {journal} {New Journal of Physics}\ }\textbf {\bibinfo {volume} {21}},\
  \bibinfo {pages} {123032} (\bibinfo {year} {2019})}\BibitemShut {NoStop}%
\bibitem [{\citenamefont {Dutton}\ \emph {et~al.}(2019)\citenamefont {Dutton},
  \citenamefont {Kerviche}, \citenamefont {Ashok},\ and\ \citenamefont
  {Guha}}]{Zachary2019}%
  \BibitemOpen
  \bibfield  {author} {\bibinfo {author} {\bibfnamefont {Z.}~\bibnamefont
  {Dutton}}, \bibinfo {author} {\bibfnamefont {R.}~\bibnamefont {Kerviche}},
  \bibinfo {author} {\bibfnamefont {A.}~\bibnamefont {Ashok}}, \ and\ \bibinfo
  {author} {\bibfnamefont {S.}~\bibnamefont {Guha}},\ }\href {\doibase
  10.1103/PhysRevA.99.033847} {\bibfield  {journal} {\bibinfo  {journal} {Phys.
  Rev. A}\ }\textbf {\bibinfo {volume} {99}},\ \bibinfo {pages} {033847}
  (\bibinfo {year} {2019})}\BibitemShut {NoStop}%
\bibitem [{\citenamefont {Prasad}(2020{\natexlab{b}})}]{Prasad2020B}%
  \BibitemOpen
  \bibfield  {author} {\bibinfo {author} {\bibfnamefont {S.}~\bibnamefont
  {Prasad}},\ }\href {\doibase 10.1103/PhysRevA.102.063719} {\bibfield
  {journal} {\bibinfo  {journal} {Phys. Rev. A}\ }\textbf {\bibinfo {volume}
  {102}},\ \bibinfo {pages} {063719} (\bibinfo {year}
  {2020}{\natexlab{b}})}\BibitemShut {NoStop}%
\bibitem [{\citenamefont {Tsang}(2019{\natexlab{b}})}]{Tsang2019a}%
  \BibitemOpen
  \bibfield  {author} {\bibinfo {author} {\bibfnamefont {M.}~\bibnamefont
  {Tsang}},\ }\href@noop {} {\bibfield  {journal} {\bibinfo  {journal}
  {Contemporary Physics}\ }\textbf {\bibinfo {volume} {60}},\ \bibinfo {pages}
  {279} (\bibinfo {year} {2019}{\natexlab{b}})}\BibitemShut {NoStop}%
\bibitem [{\citenamefont {Zhou}\ and\ \citenamefont {Jiang}(2019)}]{Zhou2019}%
  \BibitemOpen
  \bibfield  {author} {\bibinfo {author} {\bibfnamefont {S.}~\bibnamefont
  {Zhou}}\ and\ \bibinfo {author} {\bibfnamefont {L.}~\bibnamefont {Jiang}},\
  }\href {\doibase 10.1103/PhysRevA.99.013808} {\bibfield  {journal} {\bibinfo
  {journal} {Physical Review A}\ }\textbf {\bibinfo {volume} {99}},\ \bibinfo
  {pages} {013808} (\bibinfo {year} {2019})}\BibitemShut {NoStop}%
\bibitem [{\citenamefont {Lu}\ \emph {et~al.}(2018)\citenamefont {Lu},
  \citenamefont {Krovi}, \citenamefont {Nair}, \citenamefont {Guha},\ and\
  \citenamefont {Shapiro}}]{Lu2018}%
  \BibitemOpen
  \bibfield  {author} {\bibinfo {author} {\bibfnamefont {X.~M.}\ \bibnamefont
  {Lu}}, \bibinfo {author} {\bibfnamefont {H.}~\bibnamefont {Krovi}}, \bibinfo
  {author} {\bibfnamefont {R.}~\bibnamefont {Nair}}, \bibinfo {author}
  {\bibfnamefont {S.}~\bibnamefont {Guha}}, \ and\ \bibinfo {author}
  {\bibfnamefont {J.~H.}\ \bibnamefont {Shapiro}},\ }\href@noop {} {\bibfield
  {journal} {\bibinfo  {journal} {npj Quantum Information}\ }\textbf {\bibinfo
  {volume} {64}} (\bibinfo {year} {2018})}\BibitemShut {NoStop}%
\bibitem [{\citenamefont {Huang}\ and\ \citenamefont {Lupo}(2021)}]{Huang2021}%
  \BibitemOpen
  \bibfield  {author} {\bibinfo {author} {\bibfnamefont {Z.}~\bibnamefont
  {Huang}}\ and\ \bibinfo {author} {\bibfnamefont {C.}~\bibnamefont {Lupo}},\
  }\href {\doibase 10.1103/PhysRevLett.127.130502} {\bibfield  {journal}
  {\bibinfo  {journal} {Physical Review Letters}\ }\textbf {\bibinfo {volume}
  {127}},\ \bibinfo {pages} {130502} (\bibinfo {year} {2021})}\BibitemShut
  {NoStop}%
\bibitem [{\citenamefont {Grace}\ and\ \citenamefont
  {Guha}(2022{\natexlab{a}})}]{Grace2021c}%
  \BibitemOpen
  \bibfield  {author} {\bibinfo {author} {\bibfnamefont {M.~R.}\ \bibnamefont
  {Grace}}\ and\ \bibinfo {author} {\bibfnamefont {S.}~\bibnamefont {Guha}},\
  }\href {\doibase https://doi.org/10.1103/PhysRevLett.129.180502} {\bibfield
  {journal} {\bibinfo  {journal} {Physical Review Letters}\ }\textbf {\bibinfo
  {volume} {129}},\ \bibinfo {pages} {180502} (\bibinfo {year}
  {2022}{\natexlab{a}})}\BibitemShut {NoStop}%
\bibitem [{\citenamefont {Bao}\ \emph {et~al.}(2021)\citenamefont {Bao},
  \citenamefont {Choi}, \citenamefont {Aggarwal},\ and\ \citenamefont
  {Jacob}}]{Bao2021}%
  \BibitemOpen
  \bibfield  {author} {\bibinfo {author} {\bibfnamefont {F.}~\bibnamefont
  {Bao}}, \bibinfo {author} {\bibfnamefont {H.}~\bibnamefont {Choi}}, \bibinfo
  {author} {\bibfnamefont {V.}~\bibnamefont {Aggarwal}}, \ and\ \bibinfo
  {author} {\bibfnamefont {Z.}~\bibnamefont {Jacob}},\ }\href@noop {}
  {\bibfield  {journal} {\bibinfo  {journal} {Opt. Lett.}\ }\textbf {\bibinfo
  {volume} {46}},\ \bibinfo {pages} {3045} (\bibinfo {year}
  {2021})}\BibitemShut {NoStop}%
\bibitem [{\citenamefont {Matlin}\ and\ \citenamefont
  {Zipp}(2022)}]{Matlin2022}%
  \BibitemOpen
  \bibfield  {author} {\bibinfo {author} {\bibfnamefont {E.~F.}\ \bibnamefont
  {Matlin}}\ and\ \bibinfo {author} {\bibfnamefont {L.~J.}\ \bibnamefont
  {Zipp}},\ }\href {\doibase 10.1038/s41598-022-06644-3} {\bibfield  {journal}
  {\bibinfo  {journal} {Scientific Reports}\ }\textbf {\bibinfo {volume} {12}}
  (\bibinfo {year} {2022}),\ 10.1038/s41598-022-06644-3}\BibitemShut {NoStop}%
\bibitem [{\citenamefont {Lee}\ \emph {et~al.}(2022{\natexlab{a}})\citenamefont
  {Lee}, \citenamefont {Gagatsos}, \citenamefont {Guha},\ and\ \citenamefont
  {Ashok}}]{Lee2022}%
  \BibitemOpen
  \bibfield  {author} {\bibinfo {author} {\bibfnamefont {K.~K.}\ \bibnamefont
  {Lee}}, \bibinfo {author} {\bibfnamefont {C.~N.}\ \bibnamefont {Gagatsos}},
  \bibinfo {author} {\bibfnamefont {S.}~\bibnamefont {Guha}}, \ and\ \bibinfo
  {author} {\bibfnamefont {A.}~\bibnamefont {Ashok}},\ }\href {\doibase
  10.1109/JSTSP.2022.3214774} {\bibfield  {journal} {\bibinfo  {journal} {IEEE
  Journal of Selected Topics in Signal Processing}\ ,\ \bibinfo {pages} {1}}
  (\bibinfo {year} {2022}{\natexlab{a}})}\BibitemShut {NoStop}%
\bibitem [{\citenamefont {Sajjad}\ \emph {et~al.}(2021)\citenamefont {Sajjad},
  \citenamefont {Grace}, \citenamefont {Zhuang},\ and\ \citenamefont
  {Guha}}]{Sajjad2021}%
  \BibitemOpen
  \bibfield  {author} {\bibinfo {author} {\bibfnamefont {A.}~\bibnamefont
  {Sajjad}}, \bibinfo {author} {\bibfnamefont {M.~R.}\ \bibnamefont {Grace}},
  \bibinfo {author} {\bibfnamefont {Q.}~\bibnamefont {Zhuang}}, \ and\ \bibinfo
  {author} {\bibfnamefont {S.}~\bibnamefont {Guha}},\ }\href {\doibase
  10.1103/PhysRevA.104.022410} {\bibfield  {journal} {\bibinfo  {journal}
  {Phys. Rev. A}\ }\textbf {\bibinfo {volume} {104}},\ \bibinfo {pages}
  {022410} (\bibinfo {year} {2021})}\BibitemShut {NoStop}%
\bibitem [{\citenamefont {Grace}\ \emph {et~al.}(2020)\citenamefont {Grace},
  \citenamefont {Dutton}, \citenamefont {Ashok},\ and\ \citenamefont
  {Guha}}]{Grace2020c}%
  \BibitemOpen
  \bibfield  {author} {\bibinfo {author} {\bibfnamefont {M.~R.}\ \bibnamefont
  {Grace}}, \bibinfo {author} {\bibfnamefont {Z.}~\bibnamefont {Dutton}},
  \bibinfo {author} {\bibfnamefont {A.}~\bibnamefont {Ashok}}, \ and\ \bibinfo
  {author} {\bibfnamefont {S.}~\bibnamefont {Guha}},\ }\href@noop {} {\bibfield
   {journal} {\bibinfo  {journal} {Journal of the Optical Society of America
  A}\ }\textbf {\bibinfo {volume} {37}},\ \bibinfo {pages} {1288} (\bibinfo
  {year} {2020})}\BibitemShut {NoStop}%
\bibitem [{\citenamefont {Lupo}\ \emph {et~al.}(2020)\citenamefont {Lupo},
  \citenamefont {Huang},\ and\ \citenamefont {Kok}}]{Cosmo2020}%
  \BibitemOpen
  \bibfield  {author} {\bibinfo {author} {\bibfnamefont {C.}~\bibnamefont
  {Lupo}}, \bibinfo {author} {\bibfnamefont {Z.}~\bibnamefont {Huang}}, \ and\
  \bibinfo {author} {\bibfnamefont {P.}~\bibnamefont {Kok}},\ }\href {\doibase
  10.1103/PhysRevLett.124.080503} {\bibfield  {journal} {\bibinfo  {journal}
  {Phys. Rev. Lett.}\ }\textbf {\bibinfo {volume} {124}},\ \bibinfo {pages}
  {080503} (\bibinfo {year} {2020})}\BibitemShut {NoStop}%
\bibitem [{\citenamefont {Wang}\ \emph {et~al.}(2021)\citenamefont {Wang},
  \citenamefont {Zhang},\ and\ \citenamefont {Lorenz}}]{Wang2021}%
  \BibitemOpen
  \bibfield  {author} {\bibinfo {author} {\bibfnamefont {Y.}~\bibnamefont
  {Wang}}, \bibinfo {author} {\bibfnamefont {Y.}~\bibnamefont {Zhang}}, \ and\
  \bibinfo {author} {\bibfnamefont {V.~O.}\ \bibnamefont {Lorenz}},\ }\href
  {\doibase 10.1103/PhysRevA.104.022613} {\bibfield  {journal} {\bibinfo
  {journal} {Physical Review A}\ }\textbf {\bibinfo {volume} {104}},\ \bibinfo
  {pages} {1} (\bibinfo {year} {2021})}\BibitemShut {NoStop}%
\bibitem [{\citenamefont {Bojer}\ \emph {et~al.}(2022)\citenamefont {Bojer},
  \citenamefont {Huang}, \citenamefont {Karl}, \citenamefont {Richter},
  \citenamefont {Kok},\ and\ \citenamefont {von Zanthier}}]{Bojer2022}%
  \BibitemOpen
  \bibfield  {author} {\bibinfo {author} {\bibfnamefont {M.}~\bibnamefont
  {Bojer}}, \bibinfo {author} {\bibfnamefont {Z.}~\bibnamefont {Huang}},
  \bibinfo {author} {\bibfnamefont {S.}~\bibnamefont {Karl}}, \bibinfo {author}
  {\bibfnamefont {S.}~\bibnamefont {Richter}}, \bibinfo {author} {\bibfnamefont
  {P.}~\bibnamefont {Kok}}, \ and\ \bibinfo {author} {\bibfnamefont
  {J.}~\bibnamefont {von Zanthier}},\ }\href {\doibase
  10.1088/1367-2630/ac5f30} {\bibfield  {journal} {\bibinfo  {journal} {New
  Journal of Physics}\ }\textbf {\bibinfo {volume} {24}},\ \bibinfo {pages}
  {043026} (\bibinfo {year} {2022})}\BibitemShut {NoStop}%
\bibitem [{\citenamefont {Sajjad}\ \emph {et~al.}(2024)\citenamefont {Sajjad},
  \citenamefont {Grace},\ and\ \citenamefont {Guha}}]{Sajjad2023}%
  \BibitemOpen
  \bibfield  {author} {\bibinfo {author} {\bibfnamefont {A.}~\bibnamefont
  {Sajjad}}, \bibinfo {author} {\bibfnamefont {M.~R.}\ \bibnamefont {Grace}}, \
  and\ \bibinfo {author} {\bibfnamefont {S.}~\bibnamefont {Guha}},\ }\href
  {\doibase 10.1103/PhysRevResearch.6.013212} {\bibfield  {journal} {\bibinfo
  {journal} {Phys. Rev. Res.}\ }\textbf {\bibinfo {volume} {6}},\ \bibinfo
  {pages} {013212} (\bibinfo {year} {2024})}\BibitemShut {NoStop}%
\bibitem [{\citenamefont {Campbell}(1987)}]{Campbell1987}%
  \BibitemOpen
  \bibfield  {author} {\bibinfo {author} {\bibfnamefont {J.}~\bibnamefont
  {Campbell}},\ }in\ \href@noop {} {\emph {\bibinfo {booktitle} {Applied
  Geodesy}}},\ \bibinfo {editor} {edited by\ \bibinfo {editor} {\bibfnamefont
  {S.}~\bibnamefont {Turner}}}\ (\bibinfo  {publisher} {Springer Berlin
  Heidelberg},\ \bibinfo {address} {Berlin, Heidelberg},\ \bibinfo {year}
  {1987})\ pp.\ \bibinfo {pages} {67--87}\BibitemShut {NoStop}%
\bibitem [{Law(2000)}]{Lawson1999}%
  \BibitemOpen
  \href@noop {} {\emph {\bibinfo {title} {Principles of Long Baseline Stellar
  Interferometry}}}\ (\bibinfo {year} {2000})\BibitemShut {NoStop}%
\bibitem [{\citenamefont {Monnier}(2003)}]{Monnier2003}%
  \BibitemOpen
  \bibfield  {author} {\bibinfo {author} {\bibfnamefont {J.~D.}\ \bibnamefont
  {Monnier}},\ }\href {\doibase 10.1088/0034-4885/66/5/203} {\bibfield
  {journal} {\bibinfo  {journal} {Reports on Progress in Physics}\ }\textbf
  {\bibinfo {volume} {66}},\ \bibinfo {pages} {789} (\bibinfo {year}
  {2003})}\BibitemShut {NoStop}%
\bibitem [{\citenamefont {Huang}\ \emph {et~al.}(2023)\citenamefont {Huang},
  \citenamefont {Salces-Carcoba}, \citenamefont {Adhikari}, \citenamefont
  {Safavi-Naeini},\ and\ \citenamefont {Jiang}}]{Huang2023}%
  \BibitemOpen
  \bibfield  {author} {\bibinfo {author} {\bibfnamefont {Y.}~\bibnamefont
  {Huang}}, \bibinfo {author} {\bibfnamefont {F.}~\bibnamefont
  {Salces-Carcoba}}, \bibinfo {author} {\bibfnamefont {R.~X.}\ \bibnamefont
  {Adhikari}}, \bibinfo {author} {\bibfnamefont {A.~H.}\ \bibnamefont
  {Safavi-Naeini}}, \ and\ \bibinfo {author} {\bibfnamefont {L.}~\bibnamefont
  {Jiang}},\ }\href@noop {} {\  (\bibinfo {year} {2023})},\ \Eprint
  {http://arxiv.org/abs/2312.09372} {arXiv:2312.09372 [quant-ph]} \BibitemShut
  {NoStop}%
\bibitem [{\citenamefont {Gottesman}\ \emph {et~al.}(2012)\citenamefont
  {Gottesman}, \citenamefont {Jennewein},\ and\ \citenamefont
  {Croke}}]{Gottesman2012}%
  \BibitemOpen
  \bibfield  {author} {\bibinfo {author} {\bibfnamefont {D.}~\bibnamefont
  {Gottesman}}, \bibinfo {author} {\bibfnamefont {T.}~\bibnamefont
  {Jennewein}}, \ and\ \bibinfo {author} {\bibfnamefont {S.}~\bibnamefont
  {Croke}},\ }\href {\doibase 10.1103/PhysRevLett.109.070503} {\bibfield
  {journal} {\bibinfo  {journal} {Phys. Rev. Lett.}\ }\textbf {\bibinfo
  {volume} {109}},\ \bibinfo {pages} {070503} (\bibinfo {year}
  {2012})}\BibitemShut {NoStop}%
\bibitem [{\citenamefont {Khabiboulline}\ \emph
  {et~al.}(2019{\natexlab{a}})\citenamefont {Khabiboulline}, \citenamefont
  {Borregaard}, \citenamefont {De~Greve},\ and\ \citenamefont
  {Lukin}}]{Khabiboulline2018}%
  \BibitemOpen
  \bibfield  {author} {\bibinfo {author} {\bibfnamefont {E.~T.}\ \bibnamefont
  {Khabiboulline}}, \bibinfo {author} {\bibfnamefont {J.}~\bibnamefont
  {Borregaard}}, \bibinfo {author} {\bibfnamefont {K.}~\bibnamefont
  {De~Greve}}, \ and\ \bibinfo {author} {\bibfnamefont {M.~D.}\ \bibnamefont
  {Lukin}},\ }\href {\doibase 10.1103/PhysRevLett.123.070504} {\bibfield
  {journal} {\bibinfo  {journal} {Phys. Rev. Lett.}\ }\textbf {\bibinfo
  {volume} {123}},\ \bibinfo {pages} {070504} (\bibinfo {year}
  {2019}{\natexlab{a}})}\BibitemShut {NoStop}%
\bibitem [{\citenamefont {Tsang}\ \emph
  {et~al.}(2016{\natexlab{b}})\citenamefont {Tsang}, \citenamefont {Nair},\
  and\ \citenamefont {Lu}}]{Tsang2016a}%
  \BibitemOpen
  \bibfield  {author} {\bibinfo {author} {\bibfnamefont {M.}~\bibnamefont
  {Tsang}}, \bibinfo {author} {\bibfnamefont {R.}~\bibnamefont {Nair}}, \ and\
  \bibinfo {author} {\bibfnamefont {X.-M.}\ \bibnamefont {Lu}},\ }\href@noop {}
  {\  (\bibinfo {year} {2016}{\natexlab{b}})},\ \Eprint
  {http://arxiv.org/abs/1602.04655} {arXiv:1602.04655} \BibitemShut {NoStop}%
\bibitem [{\citenamefont {Grace}\ and\ \citenamefont
  {Guha}(2022{\natexlab{b}})}]{Grace2022}%
  \BibitemOpen
  \bibfield  {author} {\bibinfo {author} {\bibfnamefont {M.~R.}\ \bibnamefont
  {Grace}}\ and\ \bibinfo {author} {\bibfnamefont {S.}~\bibnamefont {Guha}},\
  }\href@noop {} {\bibfield  {journal} {\bibinfo  {journal} {Phys. Rev. Lett.}\
  }\textbf {\bibinfo {volume} {129}},\ \bibinfo {pages} {180502} (\bibinfo
  {year} {2022}{\natexlab{b}})}\BibitemShut {NoStop}%
\bibitem [{\citenamefont {Clements}\ \emph {et~al.}(2016)\citenamefont
  {Clements}, \citenamefont {Humphreys}, \citenamefont {Metcalf}, \citenamefont
  {Kolthammer},\ and\ \citenamefont {Walmsley}}]{Clements2016}%
  \BibitemOpen
  \bibfield  {author} {\bibinfo {author} {\bibfnamefont {W.~R.}\ \bibnamefont
  {Clements}}, \bibinfo {author} {\bibfnamefont {P.~C.}\ \bibnamefont
  {Humphreys}}, \bibinfo {author} {\bibfnamefont {B.~J.}\ \bibnamefont
  {Metcalf}}, \bibinfo {author} {\bibfnamefont {W.~S.}\ \bibnamefont
  {Kolthammer}}, \ and\ \bibinfo {author} {\bibfnamefont {I.~A.}\ \bibnamefont
  {Walmsley}},\ }\href {\doibase 10.1364/OPTICA.3.001460} {\bibfield  {journal}
  {\bibinfo  {journal} {Optica}\ }\textbf {\bibinfo {volume} {3}},\ \bibinfo
  {pages} {1460} (\bibinfo {year} {2016})}\BibitemShut {NoStop}%
\bibitem [{\citenamefont {Chou}\ \emph {et~al.}(2018)\citenamefont {Chou},
  \citenamefont {Blumoff}, \citenamefont {Wang}, \citenamefont {Reinhold},
  \citenamefont {Axline}, \citenamefont {Gao}, \citenamefont {Frunzio},
  \citenamefont {Devoret}, \citenamefont {Jiang},\ and\ \citenamefont
  {Schoelkopf}}]{Chou2018}%
  \BibitemOpen
  \bibfield  {author} {\bibinfo {author} {\bibfnamefont {K.~S.}\ \bibnamefont
  {Chou}}, \bibinfo {author} {\bibfnamefont {J.~Z.}\ \bibnamefont {Blumoff}},
  \bibinfo {author} {\bibfnamefont {C.~S.}\ \bibnamefont {Wang}}, \bibinfo
  {author} {\bibfnamefont {P.~C.}\ \bibnamefont {Reinhold}}, \bibinfo {author}
  {\bibfnamefont {C.~J.}\ \bibnamefont {Axline}}, \bibinfo {author}
  {\bibfnamefont {Y.~Y.}\ \bibnamefont {Gao}}, \bibinfo {author} {\bibfnamefont
  {L.}~\bibnamefont {Frunzio}}, \bibinfo {author} {\bibfnamefont {M.~H.}\
  \bibnamefont {Devoret}}, \bibinfo {author} {\bibfnamefont {L.}~\bibnamefont
  {Jiang}}, \ and\ \bibinfo {author} {\bibfnamefont {R.~J.}\ \bibnamefont
  {Schoelkopf}},\ }\href {\doibase 10.1038/s41586-018-0470-y} {\bibfield
  {journal} {\bibinfo  {journal} {Nature}\ }\textbf {\bibinfo {volume} {561}},\
  \bibinfo {pages} {368} (\bibinfo {year} {2018})}\BibitemShut {NoStop}%
\bibitem [{\citenamefont {Tang}\ \emph {et~al.}(2016)\citenamefont {Tang},
  \citenamefont {Durak},\ and\ \citenamefont {Ling}}]{Tang2016}%
  \BibitemOpen
  \bibfield  {author} {\bibinfo {author} {\bibfnamefont {Z.~S.}\ \bibnamefont
  {Tang}}, \bibinfo {author} {\bibfnamefont {K.}~\bibnamefont {Durak}}, \ and\
  \bibinfo {author} {\bibfnamefont {A.}~\bibnamefont {Ling}},\ }\href {\doibase
  10.1364/OE.24.022004} {\bibfield  {journal} {\bibinfo  {journal} {Opt.
  Express}\ }\textbf {\bibinfo {volume} {24}},\ \bibinfo {pages} {22004}
  (\bibinfo {year} {2016})}\BibitemShut {NoStop}%
\bibitem [{\citenamefont {Khabiboulline}\ \emph
  {et~al.}(2019{\natexlab{b}})\citenamefont {Khabiboulline}, \citenamefont
  {Borregaard}, \citenamefont {De~Greve},\ and\ \citenamefont
  {Lukin}}]{Khabiboulline2019}%
  \BibitemOpen
  \bibfield  {author} {\bibinfo {author} {\bibfnamefont {E.~T.}\ \bibnamefont
  {Khabiboulline}}, \bibinfo {author} {\bibfnamefont {J.}~\bibnamefont
  {Borregaard}}, \bibinfo {author} {\bibfnamefont {K.}~\bibnamefont
  {De~Greve}}, \ and\ \bibinfo {author} {\bibfnamefont {M.~D.}\ \bibnamefont
  {Lukin}},\ }\href {\doibase 10.1103/PhysRevA.100.022316} {\bibfield
  {journal} {\bibinfo  {journal} {Phys. Rev. A}\ }\textbf {\bibinfo {volume}
  {100}},\ \bibinfo {pages} {022316} (\bibinfo {year}
  {2019}{\natexlab{b}})}\BibitemShut {NoStop}%
\bibitem [{\citenamefont {Chen}\ \emph {et~al.}(2023)\citenamefont {Chen},
  \citenamefont {Nomerotski}, \citenamefont {Slosar}, \citenamefont {Stankus},\
  and\ \citenamefont {Vintskevich}}]{Chen2023}%
  \BibitemOpen
  \bibfield  {author} {\bibinfo {author} {\bibfnamefont {Z.}~\bibnamefont
  {Chen}}, \bibinfo {author} {\bibfnamefont {A.}~\bibnamefont {Nomerotski}},
  \bibinfo {author} {\bibfnamefont {A.~c.~v.}\ \bibnamefont {Slosar}}, \bibinfo
  {author} {\bibfnamefont {P.}~\bibnamefont {Stankus}}, \ and\ \bibinfo
  {author} {\bibfnamefont {S.}~\bibnamefont {Vintskevich}},\ }\href {\doibase
  10.1103/PhysRevD.107.023015} {\bibfield  {journal} {\bibinfo  {journal}
  {Phys. Rev. D}\ }\textbf {\bibinfo {volume} {107}},\ \bibinfo {pages}
  {023015} (\bibinfo {year} {2023})}\BibitemShut {NoStop}%
\bibitem [{\citenamefont {Czupryniak}\ \emph {et~al.}(2023)\citenamefont
  {Czupryniak}, \citenamefont {Steinmetz}, \citenamefont {Kwiat},\ and\
  \citenamefont {Jordan}}]{Czupryniak2023}%
  \BibitemOpen
  \bibfield  {author} {\bibinfo {author} {\bibfnamefont {R.}~\bibnamefont
  {Czupryniak}}, \bibinfo {author} {\bibfnamefont {J.}~\bibnamefont
  {Steinmetz}}, \bibinfo {author} {\bibfnamefont {P.~G.}\ \bibnamefont
  {Kwiat}}, \ and\ \bibinfo {author} {\bibfnamefont {A.~N.}\ \bibnamefont
  {Jordan}},\ }\href {\doibase 10.1103/PhysRevA.108.052408} {\bibfield
  {journal} {\bibinfo  {journal} {Phys. Rev. A}\ }\textbf {\bibinfo {volume}
  {108}},\ \bibinfo {pages} {052408} (\bibinfo {year} {2023})}\BibitemShut
  {NoStop}%
\bibitem [{\citenamefont {Pant}\ \emph {et~al.}(2017)\citenamefont {Pant},
  \citenamefont {Krovi}, \citenamefont {Englund},\ and\ \citenamefont
  {Guha}}]{Pant2017}%
  \BibitemOpen
  \bibfield  {author} {\bibinfo {author} {\bibfnamefont {M.}~\bibnamefont
  {Pant}}, \bibinfo {author} {\bibfnamefont {H.}~\bibnamefont {Krovi}},
  \bibinfo {author} {\bibfnamefont {D.}~\bibnamefont {Englund}}, \ and\
  \bibinfo {author} {\bibfnamefont {S.}~\bibnamefont {Guha}},\ }\href@noop {}
  {\bibfield  {journal} {\bibinfo  {journal} {Phys. Rev. A}\ }\textbf {\bibinfo
  {volume} {95}},\ \bibinfo {pages} {012304} (\bibinfo {year}
  {2017})}\BibitemShut {NoStop}%
\bibitem [{\citenamefont {Czupryniak}\ \emph {et~al.}(2022)\citenamefont
  {Czupryniak}, \citenamefont {Chitambar}, \citenamefont {Steinmetz},\ and\
  \citenamefont {Jordan}}]{Czu2022}%
  \BibitemOpen
  \bibfield  {author} {\bibinfo {author} {\bibfnamefont {R.}~\bibnamefont
  {Czupryniak}}, \bibinfo {author} {\bibfnamefont {E.}~\bibnamefont
  {Chitambar}}, \bibinfo {author} {\bibfnamefont {J.}~\bibnamefont
  {Steinmetz}}, \ and\ \bibinfo {author} {\bibfnamefont {A.~N.}\ \bibnamefont
  {Jordan}},\ }\href {\doibase 10.1103/PhysRevA.106.032424} {\bibfield
  {journal} {\bibinfo  {journal} {Phys. Rev. A}\ }\textbf {\bibinfo {volume}
  {106}},\ \bibinfo {pages} {032424} (\bibinfo {year} {2022})}\BibitemShut
  {NoStop}%
\bibitem [{\citenamefont {Guha}\ \emph {et~al.}(2015)\citenamefont {Guha},
  \citenamefont {Krovi}, \citenamefont {Fuchs}, \citenamefont {Dutton},
  \citenamefont {Slater},\ and\ \citenamefont {{others}}}]{Guha2015}%
  \BibitemOpen
  \bibfield  {author} {\bibinfo {author} {\bibfnamefont {S.}~\bibnamefont
  {Guha}}, \bibinfo {author} {\bibfnamefont {H.}~\bibnamefont {Krovi}},
  \bibinfo {author} {\bibfnamefont {C.~A.}\ \bibnamefont {Fuchs}}, \bibinfo
  {author} {\bibfnamefont {Z.}~\bibnamefont {Dutton}}, \bibinfo {author}
  {\bibfnamefont {J.~A.}\ \bibnamefont {Slater}}, \ and\ \bibinfo {author}
  {\bibnamefont {{others}}},\ }\href@noop {} {\bibfield  {journal} {\bibinfo
  {journal} {Phys. Rev. A}\ } (\bibinfo {year} {2015})}\BibitemShut {NoStop}%
\bibitem [{\citenamefont {Dhara}\ \emph {et~al.}(2021)\citenamefont {Dhara},
  \citenamefont {Patil}, \citenamefont {Krovi},\ and\ \citenamefont
  {Guha}}]{Dhara2021}%
  \BibitemOpen
  \bibfield  {author} {\bibinfo {author} {\bibfnamefont {P.}~\bibnamefont
  {Dhara}}, \bibinfo {author} {\bibfnamefont {A.}~\bibnamefont {Patil}},
  \bibinfo {author} {\bibfnamefont {H.}~\bibnamefont {Krovi}}, \ and\ \bibinfo
  {author} {\bibfnamefont {S.}~\bibnamefont {Guha}},\ }\href@noop {} {\bibfield
   {journal} {\bibinfo  {journal} {Phys. Rev. A}\ }\textbf {\bibinfo {volume}
  {104}},\ \bibinfo {pages} {052612} (\bibinfo {year} {2021})}\BibitemShut
  {NoStop}%
\bibitem [{\citenamefont {Dhara}\ \emph {et~al.}(2023)\citenamefont {Dhara},
  \citenamefont {Englund},\ and\ \citenamefont {Guha}}]{Dhara2023}%
  \BibitemOpen
  \bibfield  {author} {\bibinfo {author} {\bibfnamefont {P.}~\bibnamefont
  {Dhara}}, \bibinfo {author} {\bibfnamefont {D.}~\bibnamefont {Englund}}, \
  and\ \bibinfo {author} {\bibfnamefont {S.}~\bibnamefont {Guha}},\ }\href@noop
  {} {\bibfield  {journal} {\bibinfo  {journal} {Phys. Rev. Res.}\ }\textbf
  {\bibinfo {volume} {5}},\ \bibinfo {pages} {033149} (\bibinfo {year}
  {2023})}\BibitemShut {NoStop}%
\bibitem [{\citenamefont {Lee}\ \emph {et~al.}(2022{\natexlab{b}})\citenamefont
  {Lee}, \citenamefont {Bersin}, \citenamefont {Dahlberg}, \citenamefont
  {Wehner},\ and\ \citenamefont {Englund}}]{Lee2022a}%
  \BibitemOpen
  \bibfield  {author} {\bibinfo {author} {\bibfnamefont {Y.}~\bibnamefont
  {Lee}}, \bibinfo {author} {\bibfnamefont {E.}~\bibnamefont {Bersin}},
  \bibinfo {author} {\bibfnamefont {A.}~\bibnamefont {Dahlberg}}, \bibinfo
  {author} {\bibfnamefont {S.}~\bibnamefont {Wehner}}, \ and\ \bibinfo {author}
  {\bibfnamefont {D.}~\bibnamefont {Englund}},\ }\href@noop {} {\bibfield
  {journal} {\bibinfo  {journal} {npj Quantum Information}\ }\textbf {\bibinfo
  {volume} {8}},\ \bibinfo {pages} {1} (\bibinfo {year}
  {2022}{\natexlab{b}})}\BibitemShut {NoStop}%
\bibitem [{\citenamefont {Helstrom}(1976)}]{Helstrom1976}%
  \BibitemOpen
  \bibfield  {author} {\bibinfo {author} {\bibfnamefont {C.~W.}\ \bibnamefont
  {Helstrom}},\ }\href@noop {} {\emph {\bibinfo {title} {Quantum Detection and
  Estimation Theory}}}\ (\bibinfo  {publisher} {Academic Press, New York},\
  \bibinfo {year} {1976})\BibitemShut {NoStop}%
\bibitem [{\citenamefont {Yuen}\ and\ \citenamefont
  {Shapiro}(1978)}]{yuen1978}%
  \BibitemOpen
  \bibfield  {author} {\bibinfo {author} {\bibfnamefont {H.}~\bibnamefont
  {Yuen}}\ and\ \bibinfo {author} {\bibfnamefont {J.}~\bibnamefont {Shapiro}},\
  }\href {\doibase 10.1109/TIT.1978.1055958} {\bibfield  {journal} {\bibinfo
  {journal} {IEEE Transactions on Information Theory}\ }\textbf {\bibinfo
  {volume} {24}},\ \bibinfo {pages} {657} (\bibinfo {year} {1978})}\BibitemShut
  {NoStop}%
\bibitem [{\citenamefont {Rehacek}\ \emph {et~al.}(2017)\citenamefont
  {Rehacek}, \citenamefont {Pa\'{u}r}, \citenamefont {Stoklasa}, \citenamefont
  {Hradil},\ and\ \citenamefont {S\'{a}nchez-Soto}}]{Rehacek2017a}%
  \BibitemOpen
  \bibfield  {author} {\bibinfo {author} {\bibfnamefont {J.}~\bibnamefont
  {Rehacek}}, \bibinfo {author} {\bibfnamefont {M.}~\bibnamefont {Pa\'{u}r}},
  \bibinfo {author} {\bibfnamefont {B.}~\bibnamefont {Stoklasa}}, \bibinfo
  {author} {\bibfnamefont {Z.}~\bibnamefont {Hradil}}, \ and\ \bibinfo {author}
  {\bibfnamefont {L.~L.}\ \bibnamefont {S\'{a}nchez-Soto}},\ }\href {\doibase
  10.1364/OL.42.000231} {\bibfield  {journal} {\bibinfo  {journal} {Opt.
  Lett.}\ }\textbf {\bibinfo {volume} {42}},\ \bibinfo {pages} {231} (\bibinfo
  {year} {2017})}\BibitemShut {NoStop}%
\bibitem [{\citenamefont {Duan}\ and\ \citenamefont
  {Kimble}(2004)}]{DuanKimble2004}%
  \BibitemOpen
  \bibfield  {author} {\bibinfo {author} {\bibfnamefont {L.-M.}\ \bibnamefont
  {Duan}}\ and\ \bibinfo {author} {\bibfnamefont {H.~J.}\ \bibnamefont
  {Kimble}},\ }\href {\doibase 10.1103/PhysRevLett.92.127902} {\bibfield
  {journal} {\bibinfo  {journal} {Phys. Rev. Lett.}\ }\textbf {\bibinfo
  {volume} {92}},\ \bibinfo {pages} {127902} (\bibinfo {year}
  {2004})}\BibitemShut {NoStop}%
\bibitem [{\citenamefont {Pa\'{u}r}\ \emph {et~al.}(2016)\citenamefont
  {Pa\'{u}r}, \citenamefont {Stoklasa}, \citenamefont {Hradil}, \citenamefont
  {S\'{a}nchez-Soto},\ and\ \citenamefont {Rehacek}}]{Paur:16}%
  \BibitemOpen
  \bibfield  {author} {\bibinfo {author} {\bibfnamefont {M.}~\bibnamefont
  {Pa\'{u}r}}, \bibinfo {author} {\bibfnamefont {B.}~\bibnamefont {Stoklasa}},
  \bibinfo {author} {\bibfnamefont {Z.}~\bibnamefont {Hradil}}, \bibinfo
  {author} {\bibfnamefont {L.~L.}\ \bibnamefont {S\'{a}nchez-Soto}}, \ and\
  \bibinfo {author} {\bibfnamefont {J.}~\bibnamefont {Rehacek}},\ }\href
  {\doibase 10.1364/OPTICA.3.001144} {\bibfield  {journal} {\bibinfo  {journal}
  {Optica}\ }\textbf {\bibinfo {volume} {3}},\ \bibinfo {pages} {1144}
  (\bibinfo {year} {2016})}\BibitemShut {NoStop}%
\bibitem [{Note1()}]{Note1}%
  \BibitemOpen
  \bibinfo {note} {The proof that the pairwise measurement with 2-dimensional
  apertures attains the QFI is beyond the scope of this work and will be shown
  in a future publication.}\BibitemShut {Stop}%
\bibitem [{\citenamefont {Gottesman}\ and\ \citenamefont
  {Chuang}(1999)}]{gottesman1999quantum}%
  \BibitemOpen
  \bibfield  {author} {\bibinfo {author} {\bibfnamefont {D.}~\bibnamefont
  {Gottesman}}\ and\ \bibinfo {author} {\bibfnamefont {I.~L.}\ \bibnamefont
  {Chuang}},\ }\href@noop {} {\bibfield  {journal} {\bibinfo  {journal} {arXiv
  preprint quant-ph/9908010}\ } (\bibinfo {year} {1999})}\BibitemShut {NoStop}%
\end{thebibliography}
\end{document}